\documentclass[12pt,preprint]{aastex}

\newcommand{\mc}[3]{\multicolumn{#1}{#2}{#3}}
\newcommand{\kms}{$\mathrm {km\,s^{-1}}$}
\newcommand{\mum}{$\mathrm {\mu m}$}

\newcommand{\mus}{$\mathrm {\mu s}$}

\slugcomment{{\em Astrophysical Journal, 2015}}

\shorttitle{ULYSSES IN-SITU INTERSTELLAR DUST SIZE DISTRIBUTION}
\shortauthors{Kr\"uger et al.}

\begin{document}


\title{16 Years of Ulysses Interstellar Dust Measurements in the Solar System: I. Mass Distribution
and Gas-to-Dust Mass Ratio}


\author{Harald~Kr\"uger and
        Peter~Strub}
\affil{Max-Planck-Institut f\"ur Sonnensystemforschung,     
              Justus-von-Liebig-Weg 3, 37077 G\"ottingen, Germany 
              \email{krueger@mps.mpg.de}}
 \author{Eberhard~Gr\"un\altaffilmark{1}}
\affil{Max-Planck-Institut f\"ur Kernphysik, Saupfercheckweg~1, 69117 Heidelberg, Germany}
 \and      
\author{Veerle~J.~Sterken\altaffilmark{2}} 
 \affil{International Space Science Institute, Hallerstrasse 6, 3012 Bern, Switzerland}


\altaffiltext{1}{Laboratory for Atmospheric and Space Physics, University of Colorado, 
             Boulder, CO, 80303-7814, USA}
\altaffiltext{2}{Max-Planck-Institut f\"ur Kernphysik, Saupfercheckweg~1, 69117 Heidelberg, Germany}

\begin{abstract}
In the early 1990s, contemporary interstellar dust penetrating deep into the heliosphere was 
identified with the in-situ dust detector on board the Ulysses spacecraft. 
Between 1992 and the end of 2007 Ulysses monitored the interstellar dust stream. 
The interstellar grains act as tracers of the 
physical conditions in the local interstellar medium surrounding our solar system.

Earlier analyses of the Ulysses interstellar dust data  measured between 1992 and 1998
implied the existence 
of a population
of 'big' interstellar grains \citep[up to $\mathrm{10^{-13}\,kg}$;][]{landgraf2000a,frisch1999a}.
The derived gas-to-dust-mass
ratio was smaller than the one derived 
from astronomical observations, implying a concentration of interstellar dust
in the very local interstellar medium \citep{gruen2000b}. 

In this paper we analyse the entire  
 data set from 16 years of Ulysses interstellar dust measurements in 
interplanetary space. This paper concentrates on the overall mass distribution of interstellar dust.
An accompanying paper \citep{strub2015} investigates time-variable phenomena in the Ulysses 
interstellar dust data, and in a third paper
we present the results from dynamical modelling of the interstellar dust flow applied to 
Ulysses \citep{sterken2015}. We use the latest values for the interstellar
hydrogen and helium densities, the interstellar helium flow speed 
of $\mathrm{v_{ISM \infty} = 23.2\,km\,s^{-1}}$, and the ratio of radiation
pressure to gravity, $\beta$, calculated for astronomical silicates. We find a gas-to-dust-mass ratio in the 
local interstellar cloud of $R_{g/d} = 193^{+85}_{-57}$, and a dust
density of $(2.1\pm 0.6)\times 10^{-24}\,\mathrm{kg\,m^{-3}}$. 
For a higher inflow speed of $\mathrm{26\,km\,s^{-1}}$, the gas-to-dust-mass ratio is 20\% higher,
and, accordingly, the dust density is lower by the same amount. The  
gas-to-dust mass ratio derived from our new analysis is compatible with the value most recently determined from  
astronomical observations \citep{slavin2008}. We confirm earlier results that the
very local interstellar medium contains 'big' (i.e. $\mathrm{\approx 1\, \mu m}$-sized)
interstellar grains. We find a dust density in the local
interstellar medium that is a factor of three lower 
than values implied by  earlier analyses. 

\end{abstract}


\section{Introduction}

\label{sec_introduction}


The term "dust" is often considered as a synonym for dirt, which is annoying and difficult to quantify. 
Astronomers who 
observe distant objects in our Galaxy and beyond have to struggle with foreground obscuration due to  
the zodiacal light in our Solar System, and with extinction by interstellar and even intergalactic 
dust. Therefore, dust is often considered a nuisance. 

On the other hand, cosmic dust particles are involved in many astrophysical processes and play a crucial
role in the cosmic lifecycle of matter.
They trace physical and chemical processes everywhere in the Universe, ranging from 
the solar system at our doorstep as far out as high-redshift galaxies. Cosmic dust provides the surface 
for complex chemical reactions and determines the thermal, ionization and dynamical state of matter 
through its interaction with electromagnetic radiation, cosmic rays and gas particles. 
Dust is not easily controlled, it rather follows its own dynamics and disperses rapidly from its 
source. This aspect, however, has a positive side: 
 Like photons, dust particles carry information about remote 
processes through space and time, and the objects they originated from. This concept is called 
"Dust Astronomy", and modern dust observations are performed with a dust telescope on a 
dust observatory in space \citep{gruen2004}.

Interstellar dust became a topic of astrophysical research
in the early 1930s when the existence of extinction, weakening, and scattering
of starlight in the interstellar medium (ISM) was realised. At that time, astronomical 
observations provided the only information about the properties of  dust in the ISM.

With the advent of dust detectors onboard spacecraft, it became possible to 
investigate dust particles in-situ. 40 years ago, analysis of the data
obtained with the dust instruments flown on a couple of spacecraft suggested 
that  contemporary interstellar dust grains can cross the heliospheric boundary and penetrate deeply 
into the heliosphere \citep{bertaux1976,wolf1976}. 

In the 1990s, this 
was undoubtedly demonstrated: the dust detector onboard the Ulysses spacecraft, which 
measured mass, speed and approach direction of the 
impacting grains, identified interstellar dust grains with radius
above $\mathrm{0.1\,\mu m}$ that were flowing through the heliosphere 
\citep{gruen1993a,gruen1994a,gruen1995a}. 
 These grains originated from the local interstellar cloud (LIC) which is the 
interstellar cloud surrounding our solar system. We follow the notation by \citet{frisch1999a}.
Details of the local interstellar setting of our solar system were also given by
\citet{redfield2008}.
 The Ulysses measurements offered the opportunity to 
probe dust from the local interstellar cloud.

The Ulysses interstellar dust measurements
were later confirmed by the Galileo \citep{baguhl1996,altobelli2005a} and Cassini spacecraft 
\citep{altobelli2003,altobelli2007,altobelli2015}, 
and interstellar impactors were also identified in the Helios dust data \citep{altobelli2006}. 
In 2006, the Stardust mission  
successfully brought a sample of collected interstellar grains to Earth \citep{westphal2014b}.
Finally, there are recent claims of detections of interstellar grains with radio 
and plasma wave instruments \citep{belheouane2012}.

Measurements of interstellar dust inside the planetary system now provide a new
window for the study of diffuse interstellar matter at our doorstep. However, the 
interstellar dust stream in the heliosphere is strongly modified from the undisturbed flow outside the
heliosphere, in particular by solar radiation effects and the Lorentz force. These modifications
have to be taken into account for a proper interpolation of the interstellar dust properties
to the interstellar medium outside the heliosphere where these grains originate from.

In addition to interstellar dust, various populations of dust originating from sources inside
the solar system were investigated  
in interplanetary space with the Ulysses and Galileo dust experiments: 
the interplanetary dust complex which is constantly replenished
by dust from asteroids and comets \citep{gruen1997a}, 
including $\beta$-meteoroids \citep[i.e. dust particles which leave the solar system on 
unbound orbits due to acceleration by radiation pressure;][]{hamilton1996a,wehry2004}, 
and dust stream particles expelled from the Jovian system by 
electromagnetic forces \citep{gruen1998}, to name only the most significant dust types 
studied so far. 
For a summary of the Ulysses dust investigations in interplanetary space see
 \citet[][references therein]{krueger2010b}. See also \citet{krueger2009a,mann2010,frisch2011,frisch2013a} 
for recent reviews of measurements and modelling of interstellar dust in the heliosphere and beyond. 

\subsection{Interstellar Dust Entering the Heliosphere}

The Ulysses in-situ dust measurements obtained in the 1990s showed that the motion of 
interstellar grains 
through the solar system is -- within the dust measurement accuracy -- parallel to 
the flow of neutral interstellar hydrogen and helium gas.  
A speed of  $\mathrm{ 26\,km\,s^{-1}}$ was adopted in these earlier analyses 
\citep{gruen1994a,baguhl1995a,witte1996,witte2004a}. 
The grains which originated from the very local interstellar environment of
our solar system  
were identified by their impact direction and impact 
speed, the latter being compatible with particles moving 
on hyperbolic heliocentric trajectories \citep{gruen1994a}. Their
dynamics depend on the grain size and is strongly affected by the 
interaction with the interplanetary magnetic field and by solar 
radiation pressure 
\citep{landgraf1999a,landgraf2000b,mann2000a,czechowski2003a,czechowski2003b,
landgraf2003,sterken2012a,sterken2013a,sterken2015}. Strong filtration of small grains 
due to electromagnetic 
forces also occurs at the heliospheric boundary \citep{linde2000}, leading to a strong 
modification of the size distribution and fluxes of grains measured inside the 
heliosphere. The interstellar dust flux modulation due to 
grain interaction with the interplanetary magnetic field during solar minimum could 
be well explained by numerical simulations \citep{landgraf1998a,landgraf2000b,landgraf2003}. 

The interstellar dust flow persists at high ecliptic latitudes above and below
the ecliptic plane and even over the poles of the Sun, whereas 
interplanetary dust is strongly depleted at high latitudes 
\citep{gruen1997a}.  The interstellar dust flux measured at a distance of 
about 3~AU from the Sun is time-dependent, and the mean mass of the grains
is about $\mathrm{3\times 10^{-16}\,kg}$ \citep{landgraf2000a}, corresponding
to a grain radius of approximately $\mathrm{0.3\mu m}$ (assuming a grain density 
of $\mathrm{2.5\,kg\,m^{-3}}$). 
The earlier analyses of the Ulysses dust measurements yielded an upstream direction 
of the dust flow  at $\mathrm{259^{\circ}}$ ecliptic
longitude and $\mathrm{8^{\circ}}$ latitude \citep{landgraf1998a}.

Spectroscopic observations of sightlines to stars enable information 
of intervening dust characteristics to be obtained. 
Studies of the dust impacts detected with both Ulysses and its twin dust detector 
on board Galileo indicated that the intrinsic size distribution of interstellar grains in 
the local interstellar cloud 
extends to grain sizes larger than those detectable by such astronomical observations
\citep{frisch1999a,frisch2003a,landgraf2000a,gruen2000b}.
Observations of radar meteors entering the Earth's atmosphere at high speeds  
indicate the existence of even larger interstellar
grains \citep{taylor1996b,baggaley2002,baggaley2007}, although this conclusion remains under
debate. 

The Ulysses and Galileo interstellar dust measurements implied that the 
gas-to-dust mass ratio in the local interstellar cloud 
is higher than the standard interstellar value derived from 
cosmic abundances \citep{landgraf1998a,frisch1999a,kimura2003a}. This implied the existence of inhomogeneities 
in the diffuse interstellar medium on relatively small length scales \citep[$\mathrm{\ll 1\,kpc}$;][]{gruen2000b}.

Due to its unique highly inclined heliocentric trajectory and very long mission duration,
Ulysses was able to monitor the 
interstellar dust flow through the solar sytem over 16 years. This time period covers 
more than two and a half revolutions of the spacecraft about the
Sun through more than 2/3 of a complete 22-year (magnetic) solar cycle 
(Figure~\ref{fig_orbitplot}). Thus,
Ulysses measured interstellar dust during solar minimum and 
solar maximum conditions of the interplanetary magnetic field (IMF).

Earlier comprehensive investigations of the interstellar impactors were mostly 
performed in the late 1990s
and relied upon the significantly smaller data set available at
the time. Until the end of the Ulysses mission, 
the interstellar dust data set has grown by more than a factor of
two so that a complete re-analysis is 
worthwhile and can give new insights into, e.g., the grain dynamics inside the heliosphere 
and into the conditions in the local interstellar environment where these grains originate. 

Recent measurements with the Interstellar Boundary Explorer (IBEX) spacecraft led to a 
revision of the interstellar gas flow vector (speed and direction) derived earlier from Ulysses measurements 
\citep[][inflow speed $\mathrm{v_{ISM \infty} = 26\,km\,s^{-1}}$]{witte2004a}. The IBEX measurements of the 
interstellar flow are also more consistent with newer and independent astronomical
measurements \citep{redfield2008}. IBEX showed that the Sun is still located within
the local interstellar cloud. The inflow speed of the interstellar medium as derived 
by IBEX is $\mathrm{v_{ISM \infty} = 23.2\,km\,s^{-1}}$ 
and the downstream flow direction is $\mathrm{l_{ISM \infty} = 79^{\circ}}$ ecliptic 
longitude and $\mathrm{b_{ISM \infty} = -5^{\circ}}$ ecliptic latitude \citep{mccomas2012}. 
Given that the impact speed of the dust grains 
affects the mass calibration of our interstellar dust measurements, we analyse the Ulysses
data in view of this reduced inflow speed. However, this speed was, like the direction, under
debate 
\citep{lallement2014,wood2015,mccomas2015}. 
The higher inflow speed of 
$\mathrm{v_{ISM \infty} = 26\,km\,s^{-1}}$ 
increases our derived gas-to-dust mass ratio by about 20\% ({\em cf.} Section~\ref{sec_discussion}).

This is the first in a series of three papers dedicated to the analysis of the full
Ulysses data set of 16 years 
of interstellar dust measurements in the heliosphere. In this paper we review the mass
distribution of interstellar grains detected in the heliosphere. Temporal variations in 
dust flux, impact direction and grain size during this time period 
are investigated by \citet{strub2011,strub2015}, and results from modelling of grain dynamics 
in the context of the observations are presented by \citet{sterken2015}. 
In Section~\ref{sec_ulysses} we briefly describe the Ulysses mission, the Ulysses dust detector 
and its operation. 
In Section~\ref{sec_identification} we derive the Ulysses interstellar dust data set, and in 
Section~\ref{sec_results} we obtain the mass distribution of interstellar grains and the
gas-to-dust mass ratio in the local interstellar cloud.
Section~\ref{sec_discussion} is a discussion and in Section~\ref{sec_conclusions} we summarise 
our conclusions.


\section{The Ulysses Dust Instrument}

\label{sec_ulysses}

\begin{figure}[ht]
   \centering
   \parbox{15cm}{
   \vspace{-0.5cm}
      \hspace{-1cm}
   \includegraphics[width=1.1\textwidth]{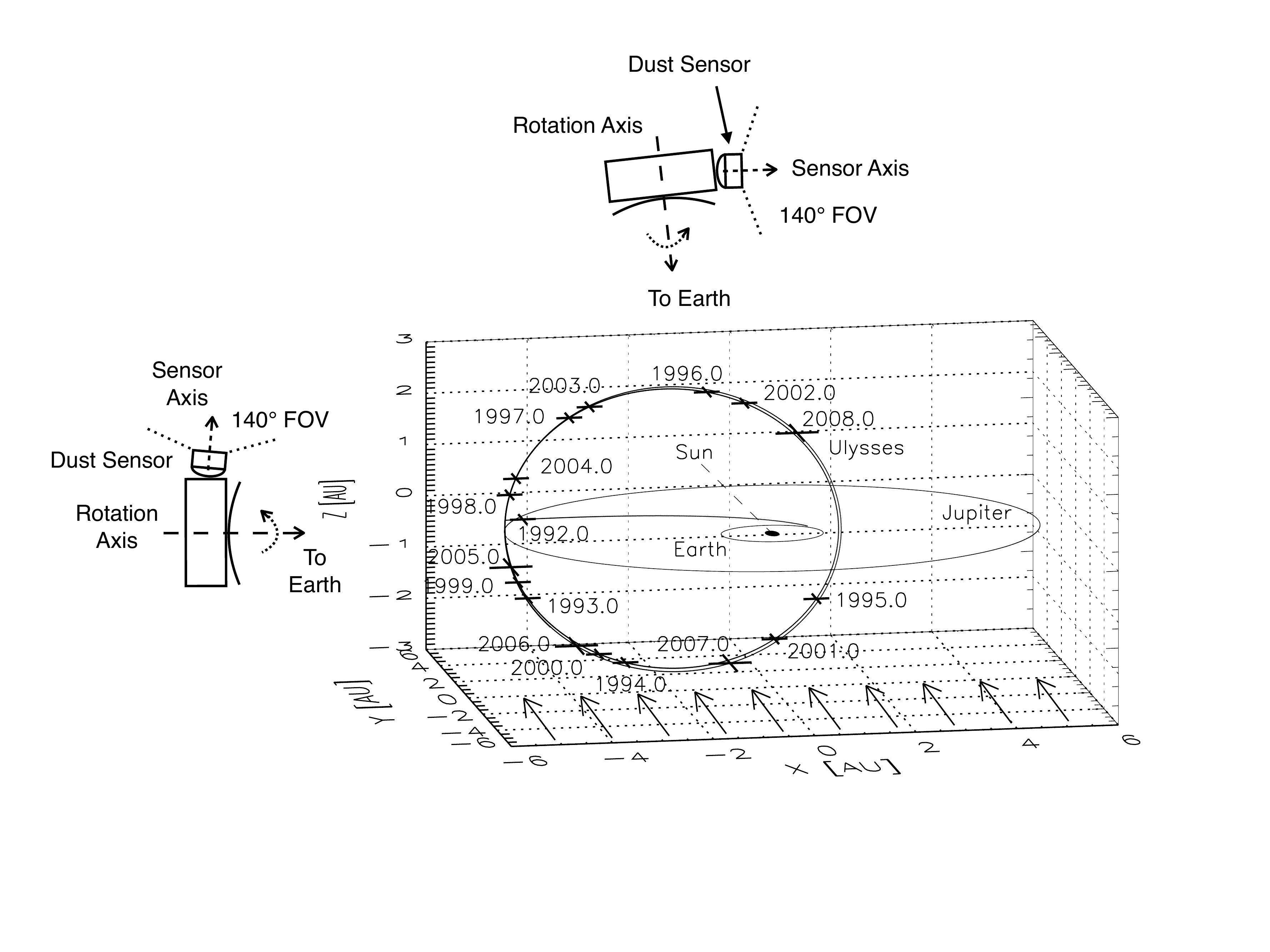}
   }
      \caption{The trajectory of Ulysses in ecliptic coordinates with the Sun at the centre. The orbits of Earth and Jupiter indicate the ecliptic plane, and the initial trajectory of Ulysses was in this plane. After Jupiter flyby in early 1992 the orbit was  almost perpendicular to the ecliptic plane ($79^{\circ}$ inclination). Crosses mark the spacecraft position at the beginning of each year. Vernal equinox is to the right (positive x axis). Arrows indicate the undisturbed interstellar dust flow direction which is within the measurement accuracy co-aligned with the direction of the interstellar helium gas flow. It is almost perpendicular to the orbital plane of Ulysses. The Ulysses spacecraft and the scan orientation of the dust detector are
      sketched for two positions along the orbit: at aphelion and at the spacecraft's highest ecliptic latitude.}
\label{fig_orbitplot}
   \end{figure}

The Ulysses dust instrument 
detects individual dust particles impacting onto the sensor target,
measures their mass and impact speed, and determines the impact
direction \citep{gruen1992b}. Up to now Ulysses was  the only space probe  that left the ecliptic plane
and passed over the poles of the Sun. Ulysses was launched 
in October 1990. After a swing-by manoeuvre at Jupiter in February 1992,
the spacecraft's orbital plane was almost perpendicular to the ecliptic plane 
($79^{\circ}$ inclination) with 
an aphelion at Jupiter and a six-year period (Figure~\ref{fig_orbitplot}). Subsequent aphelion 
passages occurred in April 1998 and in June 2004. This special orbit orientation 
allowed the dust detector on board Ulysses to 
unambiguously detect interstellar dust grains entering the heliosphere because the spacecraft's 
orbital plane was almost perpendicular to the flow
direction of the interstellar dust (Figure~\ref{fig_orbitplot}). Ulysses was operated until 2009.

A practically identical twin instrument 
was operated on board the Galileo spacecraft, which was launched 
in 1989, and between 1995 and 2003 it was the first Jupiter-orbiting spacecraft \citep{gruen1992a}. 
A third identical instrument (GORID), an engineering model
of the Ulysses sensor, was operational in geostationary orbit 
on the Express telecommunication satellite 
between 1997 and 2002 \citep{drolshagen1999}. Finally, the Cassini spacecraft,
launched in 1997, carries the Cosmic Dust Analyzer (CDA) which is 
an upgrade of the Ulysses instrument that is equipped with a time-of-flight mass spectrometer \citep{srama2004}. 
Cassini has been successfully measuring dust in the Saturnian system since 2004. 
Altogether, these
four instruments very successfully collected cosmic dust measurements
during more than 50 years in space. 

\subsection{Impact Ionisation}   
 
\label{sec_impact_ionisation}

The physical mechanism most generally utilised in modern spaceborne 
in-situ detectors of cosmic dust is based on the measurement
of the electric charge generated upon impact of a fast 
projectile onto a solid target \citep[impact ionisation,][]{raizer1960,friichtenicht1963}. 
It yields the highest 
sensitivity for the detection of dust particles in space
\citep[see][for a review]{auer2001}. The electrical charge generated upon particle impact 
can be quantitatively calibrated to provide impact speed and mass of the 
grains. The impacts can be detected by several independent measurements on different 
instrument channels (multi-coincidence detection) which allows for a reliable dust
impact detection and identification of noise events. In combination with a time-of-flight 
mass spectrometer an impact ionisation detector can measure the chemical composition of the 
impacting grains. 

When a dust particle strikes a solid target with high speed ($\gg \mathrm{1\,km\,s^{-1}}$),
it produces a crater in the target and ejecta composed of both particle
and target material. The ejecta consist of positive and negative ions, 
electrons, and neutral atoms and molecules originating from 
both projectile and target. Because of its high internal pressure, 
(up to 5 TPa) the ejecta cloud expands rapidly into the surrounding vacuum. 
As the ejecta strike sensor side walls and other surfaces, they
produce secondary ions, electrons and debris which, in turn, can
strike more surfaces and produce additional ejecta.

The experimental arrangement typically consists of a metal target plate 
and a collector (e.g. a metal grid) for either the ions or 
electrons of the impact plasma. The target is 
preferentially made of a material with a high electron
yield like molybdenum, tantalum, tungsten or gold. Different
electric potentials applied to the target plate and the
collector generate an electric field, separating the
positively and negatively charged particles of the plasma. 
Charge-sensitive amplifiers coupled to both the target
plate and the collector register independently, but 
simultaneously, an impacting dust particle. The total
amount of charge, $Q$, collected on each channel
is a function of mass $m$ and impact speed $v$ of the 
particle as well as the particle's composition. $Q$ 
can be described by the empirical law 
\begin{equation}
   Q \varpropto m^{\alpha}\, v^{\gamma} ,         \label{eq_charge}
\end{equation}
with $\alpha \simeq 1$ and $ 1.5 \lesssim \gamma  \lesssim 5.5$ in
the speed range 2 \kms\ $\lesssim v \lesssim $ 70 \kms\ 
\citep{auer2001}. In particular, for constant impact
speed, the charge generated upon impact
is proportional to the particle mass \citep{goeller1985}.

The rather wide range in $\gamma$ 
is due to different impact speeds,
target and projectile materials and collector geometries used 
for the measurements. In particular, the physical processes 
involved are speed dependent and 
the impact ionisation process is often divided into three 
speed regimes, characterised by different values of $\gamma$
\citep{stuebig2002}. 
At speeds below about 6\,\kms\ surface ionisation dominates
($3.5 \lesssim \gamma \lesssim 4.5$): The 
surfaces of the solid bodies involved in the impact process are
heated by the impact shock, leading to thermal ionisation of the
surfaces. In addition, ionisation of 
alkali contaminants on the target, having low ionisation potentials,
takes place. In the high impact speed regime, above 18\,\kms, 
target and projectile ionisation (volume ionisation) dominate 
($3.0 \lesssim \gamma \lesssim 5.5$). For 
intermediate speeds the charge yield is 
reduced due to energy consumption by melting and vaporisation 
processes ($1.5 \lesssim \gamma \lesssim 2.5$). 

A similar behaviour
was also reported by \citet{goeller1989}. Due to the large 
uncertainties in the exponent and material dependencies,
these authors used a value of $\gamma = 3.5$ throughout the 
entire speed range to calibrate the Ulysses 
dust instrument. Here we use the empirical calibration curve of 
\citet[][their Table~4c]{gruen1995a} that was derived from impact experiments at a dust 
accelerator with iron, zinc coated silica and carbon particles. 

The second important variable for the determination of the
impact parameters is the rise time of the charge signal. 
It depends only on the impact speed of the particles 
\citep{dietzel1973}. The rise time can be used to determine
the impact speed when it is in the range of
1 \kms\ $ \lesssim  v \lesssim $ 20 \kms. 

In addition to mass and speed, 
the composition of the ions in the plasma cloud can be determined
with a time-of-flight mass spectrometer separating the ions
according to their mass. 
The Ulysses  dust instrument does not have a time-of-flight
mass spectrometer, contrary to the dust instrument on board
Cassini \citep{srama2004} which is an upgrade of the 
Ulysses instrument.

\subsection{Instrument Description}

\label{sec_instrument_description}

\begin{figure}
\vspace{-6cm}
\parbox{0.549\hsize}{
\hspace{-2.4cm}
   \includegraphics[width=0.83\textwidth]{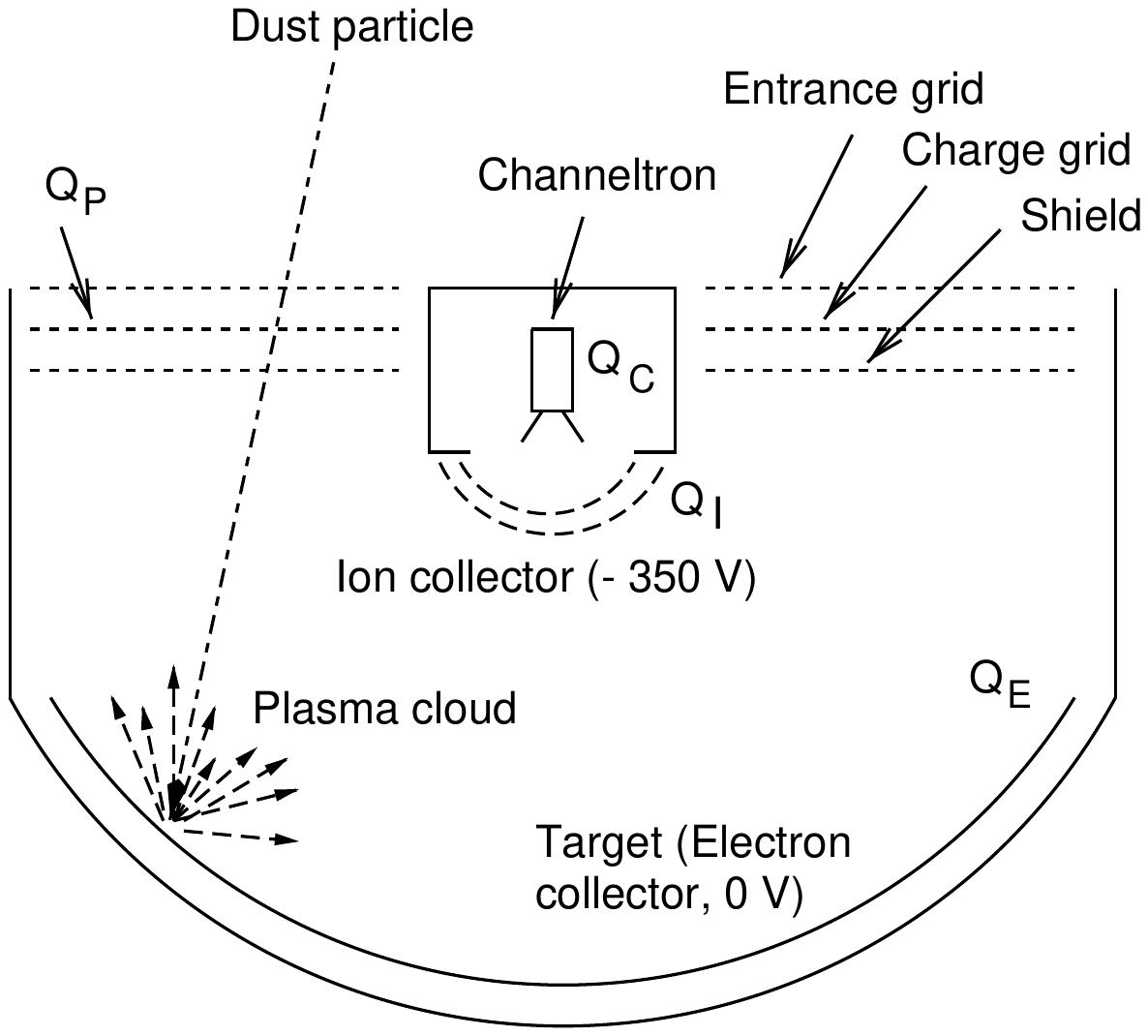}
}
\parbox{0.459\hsize}{
\hspace{-1.5cm}
   \includegraphics[width=0.63\textwidth]{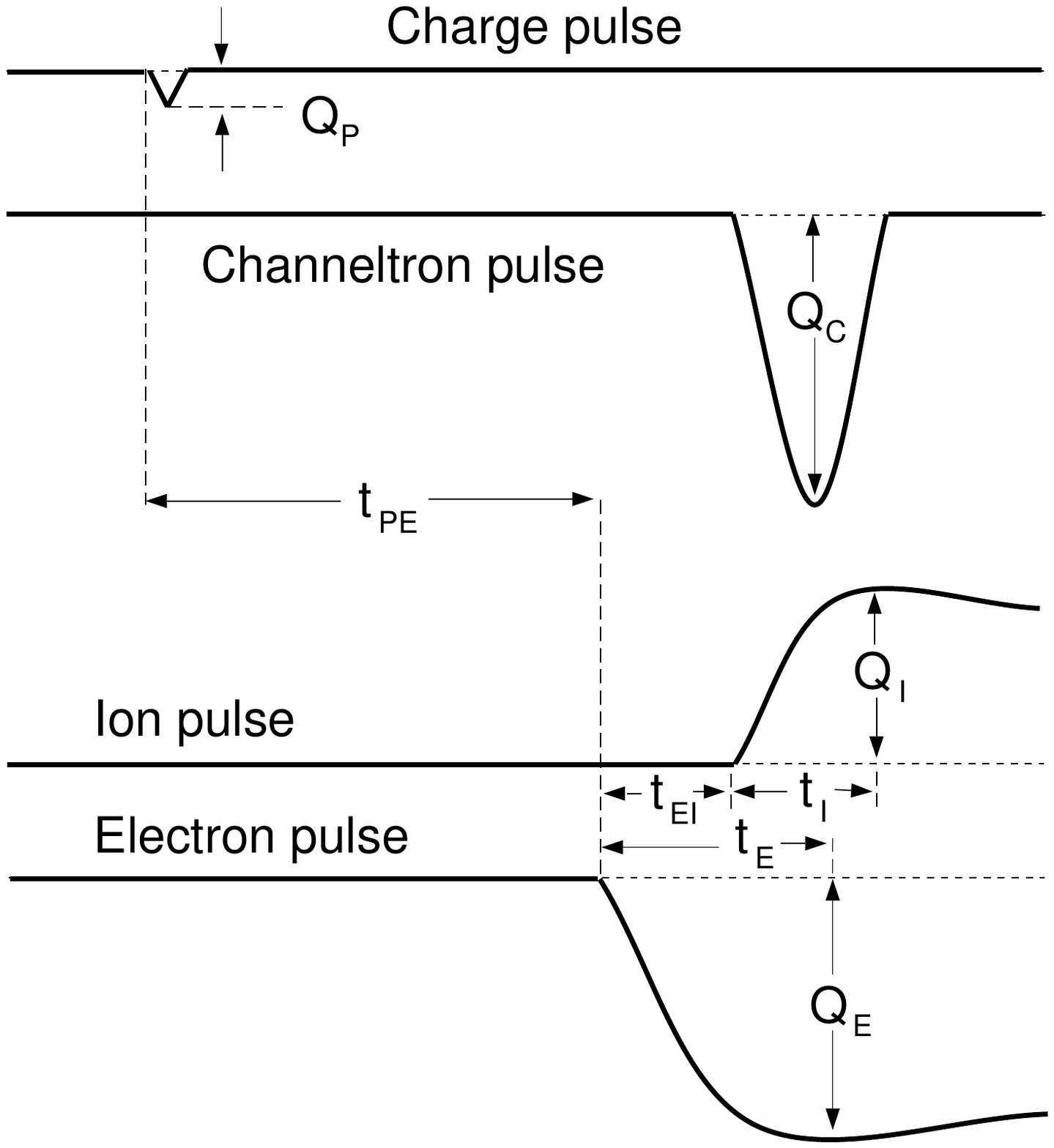}
}
\vspace{-5cm}
\caption[Dust detector configuration and measured charge signals]
{Schematic sensor configuration of the Ulysses dust detector 
({left}) and charge signals measured upon impact of a negatively 
charged dust particle ({right}).
From \citet{gruen1992a}.
}
\label{fig_sensor}
\end{figure}

The dust instrument on board Ulysses 
consists of a cylindrical sensor (with diameter 442\,mm and length 301\,mm)
with channeltron and pre-amplifiers, signal conditioning, and spacecraft
interface electronics. The sensor 
and the charge signals measured upon impact of a dust particle are schematically 
shown in Figure~\ref{fig_sensor}. 

The sensor consists of a grid system for the measurement of the particle charge,
an electrically grounded hemispherical gold-coated metal target and a negatively 
biased ion
collector grid. A charged dust particle entering the sensor induces a charge 
in the charge grid which is measured by a charge sensitive amplifier. Once the
particle hits the target, it generates electrons and ions which are separated by
the electric field of $ \rm - 350\,V$ between the hemisphere and the ion 
collector. The negative charges (electrons and negative ions, $Q_E$) are 
collected at the hemisphere and measured by a charge sensitive amplifier. 
Positive ions ($Q_I$) are collected and measured at the negatively biased ion 
collector with a charge sensitive amplifier. The ion collector has a transparency
of about 40\% so that some of the ions can penetrate the ion collector, are 
further accelerated and detected by an electron multiplier (channeltron). 
Secondary electrons are produced in the channeltron, amplified, and measured
by a charge sensitive amplifier ($Q_C$). Other parameters measured upon impact
are the rise times $t_I$ and $t_E$ of both the positive and the negative 
charge pulses $Q_I$ and $Q_E$. The measured time delay $t_{EI}$ 
between the electron and ion pulses is used to distinguish true dust 
impacts from noise events \citep{baguhl1993a}. 
Dust impacts have time delays of $2-44$ \mus,
while mechanical noise has a time delay of milliseconds.
The thresholds and dynamic ranges of the 
various signals measured upon impact are given in 
Table~\ref{tab_parameters}. 

A measurement cycle of the instrument can be initiated if one or more of the signals 
$Q_E$, $Q_I$ or $Q_C$ exceeds an adjustable threshold. During normal
operation, an event is initiated by the signals $Q_I$ or $Q_C$. 
Because
of high noise rates encountered for the electron channel, $Q_E$, this 
channel  was not selected to initiate a measurement cycle.

The parameters of a single recorded event listed in Table~\ref{tab_parameters} 
are digitised and stored in an Experiment Data Frame. Coincidences between various
event signals, event time and sensor pointing direction during the event 
as well as status information (housekeeping data) are also
recorded for each event. These data are transmitted to Earth and, in an
initial step, used to determine whether the event was a true dust 
impact or a noise event. If the measured signals were due to a dust 
impact, the particle mass $m$, and impact speed $v$ 
are derived from the instrument calibration. No instrinsic dust
charges were derived from the measured $Q_P$ signals
(Section~\ref{sec_calibration}). 
More detailed descriptions of the 
dust instrument, the reduction of the Ulysses dust data and the 
identification of noise events are given by \citet{gruen1992a,gruen1992b,gruen1995a} 
and \citet{baguhl1993a}.

\subsection{Instrument Calibration}

\label{sec_calibration}

Before an instrument is carried into space, it must be tested on the
ground to verify and calibrate its response. The most striking
characteristic of dust particles detected in space is their high
speed which is typically in the range of 1 -- 100~\kms. Their
sizes are in the range 0.01 -- 10 \mum. Thus, in order to calibrate 
a dust instrument to be flown on a space mission, one has to 
accelerate particles in this size range to comparable speeds.

The only technique with which this speed and
mass range is accessible is electrostatic acceleration
\citep{fechtig1978,auer2001}. This technique is based on the 
acquisition of kinetic energy by a particle of mass $m$ and positive
charge $q$ falling through a potential difference $U$: 
$
\frac{1}{2} m v^2 = qU           
$,
where $v$ is the terminal speed of the particle. Since the 
acceleration voltage can easily be measured, and $v$ and $q$ can
be measured with pick-up electrodes, the mass can be calculated for
each accelerated particle. 

In an electrostatic accelerator only conducting particles 
can be accelerated. Either the particle material must be a conductor,
or the particle must be coated with a conducting material.
Materials used for the calibration of the Ulysses 
detector were iron, carbon and zinc coated silica
\citep{goeller1989}. Calibration experiments for the 
Cassini dust instrument were also performed with coated latex 
particles \citep{stuebig2001,stuebig2002}.  
Recently, metal and polymer coated particles could also be used 
for calibration experiments \citep{hillier2009,hillier2014},  
and first shots with porous particle analogues 
have been and are currently being attempted \citep{sterken2013b,sterken2015}.

The calibration experiments of the Ulysses dust detector 
were performed at the Heidelberg dust accelerator
facility \citep{goeller1985,goeller1989} which is an electrostatic
accelerator with a 2 MV van de Graaf high voltage generator. The
particles were in the speed range 1\,\kms  $ \leq v \leq$ 70 \kms\ and in the 
mass range $\mathrm{10^{-18}\,kg} \leq m \leq \mathrm{10^{-13}\,kg}$. 
In addition to three different particle materials, tests with 
varying impact angles were also performed. 

\subsubsection{Speed and Mass}

The particle speed can be determined from the rise times $t_I$ and 
$t_E$ of the charge signals measured on the ion collector and on the
target. \citet{gruen1995a} measured the rise times of the
impact signals as a function of impact speed for three different
materials. The signal strength depends moderately on the particle 
material and also on the impact angle. Neither the
particle material nor the impact angle are known for an impinging
micrometeoroid. Therefore, averaged calibration curves are used to 
obtain impact speeds, assuming that the materials used for 
calibration represent cosmic dust particles of 
either iron, rock, carbonaceous or CHON composition.
Since the two rise times are measured independently, one
obtains two (often different) speed values, $v_{t_I}$ and
$v_{t_E}$. The impact speed is taken as the geometric mean
of both values: 
$
v=\sqrt{v_{t_I}\cdot v_{t_E}}.
$
The typical accuracy of the derived speed $v$ is a factor of 2.

Once the particle speed has been determined, its charge to 
mass ratio can be derived from the calibration curves obtained by
laboratory impact experiments \citep{gruen1995a}. 
From these values and the
corresponding impact charges $Q_E$ and $Q_I$, two 
independent estimates of the mass $m_{Q_E}$ and $m_{Q_I}$ 
are derived. The particle mass is usually taken as the geometric mean 
$m = \sqrt{m_{Q_E} \cdot m_{Q_I}}$.
If the speed is well determined, the mass can also be derived with a
higher accuracy. The typical uncertainty in the mass
$m$ is a factor of 10.  In this paper we derive the grain mass only from 
the charge $Q_I$ measured on the ion collector.

\begin{figure}
   \includegraphics[width=1.\textwidth]{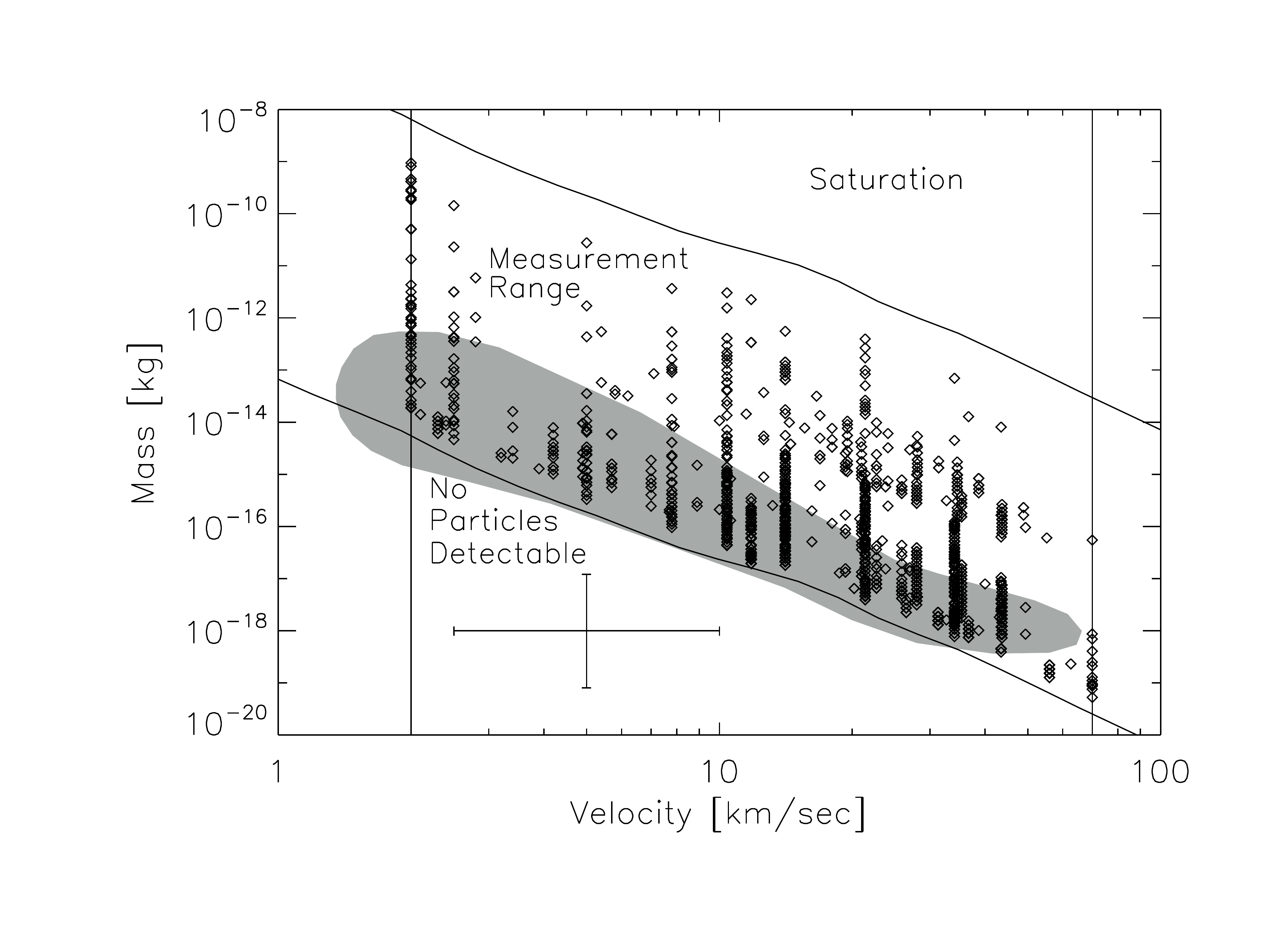}
\caption[Calibrated speed and mass range for dust detection] 
{Calibrated mass and speed range of the Ulysses
dust detector. In the region marked `Saturation' the instrument
operates as a threshold detector. The shaded area shows the 
range where the instrument was calibrated in the laboratory.
Below 2~\kms\ and above 
70~\kms\ speeds and masses cannot be determined. The 
bottom cross represents typical accuracies of speed and
mass values. Plus signs show the calibrated
masses and speeds of 2113 particles measured with Ulysses.
Jupiter stream particles are not shown as they are actually smaller 
and faster than the calibrated range of the instrument.
Adapted from \citet{gruen1992a}.
}
\label{fig_mass_speed}
\end{figure}

The speed dependent measurable mass range of the instrument
is shown in Figure~\ref{fig_mass_speed}. During the close Jupiter flyby in 1992 
the electronic detection threshold was set to a higher value because an
increased noise level was expected \citep{gruen1995c}. In this case the 
sensitivity was reduced. 

Since the charge sensitive 
amplifiers covered six orders of magnitude in impact charge (and so
did the mass range), the upper limit of the calibrated 
range is also indicated in Figure~\ref{fig_mass_speed}. For larger particles the instrument 
operated as a threshold detector (saturation range). 
The calibration covered a speed interval 
2~\kms\ $\lesssim v \lesssim$ 70~\kms.
Due to the speed-dependent mass threshold the 
mass range accessible by the instrument was 
$\rm 10^{-19}\, kg \lesssim$ $ m \lesssim  10^{-9} \rm \,kg$. 

The instrument calibration obtained in the laboratory
was confirmed by measurements in the close 
vicinity of Jupiter's Galilean moons \citep{krueger1999d}.
The moons are surrounded by clouds of 
ejecta dust grains kicked-up from their surfaces.
The average impact speeds of these, most likely
icy, grains were 6 -- 8 \kms\ and
were very close to the expected speeds.
Particle sizes were 0.5 -- 1.0 \mum\ 
\citep{krueger2000a,krueger2003b}, which
is within the well calibrated range of the instrument.
On the other hand, the jovian
dust streams first detected in interplanetary space and
later extensively studied in the jovian 
magnetosphere consist of much smaller and faster particles,
far beyond the calibrated range of the 
instrument \citep{zook1996}: grain radii were actually about 10\,nm and their speeds exceeded $\mathrm{200\,km\,s^{-1}}$.
These particles strongly interact with the interplanetary and the Jovian magnetic fields
\citep{gruen1998,flandes2011}, and they
originate from Jupiter's moon Io \citep{graps2000a}. They are
not the subject of this paper. 

\subsubsection{Charge}

The induced charge signal, $Q_P$, is a measure of the intrinsic
charge carried by a dust particle entering the dust sensor.
For two reasons the induced charge measurement is the most 
difficult measurement of the dust instrument: 
1) Cosmic dust particles are only weakly charged.
A surface potential $U$ results in a dust charge 
$ q = 4 \pi \epsilon_0 U s $
for a spherical particle with radius $s$ 
($\epsilon_0= 8.854\times \rm 10^{-12}\,A\,s\,V^{-1}\,m^{-1}$).
For a typical potential of $U = \rm 5\,V$, the smallest particle 
exceeding the detection threshold has a radius of about 20 \mum,
or, assuming a density of $\mathrm{3.3 \times 10^3\,kg\,m^{-3}}$, a corresponding 
mass of about $\mathrm{10^{-10}\,kg}$. 
The majority of the particles detected during the Ulysses 
mission had masses below $\rm 10^{-12}\,kg$ 
(Figure~\ref{fig_mass_speed}).
2) The charge grid is the measuring channel most exposed to 
ambient noise. Thus, analysis of the charge measurements 
requires careful consideration of the noise. As a consequence,
no charges have yet been determined from the Ulysses 
and Galileo dust data \citep{svestka1996}. 

The Cassini dust instrument is by an order of magnitude more sensitive in $Q_P$
which led to the detection of 
the intrinsic charges for several interplanetary particles and
for particles in Saturn's E ring \citep{kempf2004,kempf2006a}.

\subsection{Angular Sensitivity and Sensor Pointing}

\label{sec_angular_sensitivity}

Ulysses was a spinning spacecraft with a period of five revolutions per minute.
The spin axis was the centre line of the craft's high-gain antenna which normally 
pointed at Earth, and most of 
the time the spin axis pointing was within 
$1^{\circ}$ of the nominal Earth direction for data transmission. This small deviation 
is usually negligible for the analysis of measurements with the 
dust detector. 
The Ulysses spacecraft and mission were explained in more detail by 
\citet{wenzel1992}. 

The Ulysses dust sensor had a 140$^{\circ}$ wide field-of-view with a sensor
area of $\rm 1000\,cm^2$ and it was mounted
on the spacecraft nearly at right angles (85$^{\circ}$) to the antenna
axis (spacecraft spin axis). Due to this mounting geometry, the
sensor was most sensitive to particles approaching from the plane
perpendicular to the spacecraft-Earth direction. The detection geometry of the sensor 
is illustrated in Figure~\ref{fig_orbitplot}. The impact direction of
dust particles was measured by the rotation angle, $\theta$, which was the sensor
viewing direction at the time of a dust impact. During one spin
revolution of the spacecraft, the rotation angle scanned through a complete
circle of 360$^{\circ}$. It was measured in a right-handed system
and $\theta = 0^{\circ}$ was defined to be the
direction closest to ecliptic north. At $\theta = 90^{\circ}$ and $270^{\circ}$ the sensor 
axis pointed nearly along the ecliptic plane. When Ulysses was at high ecliptic latitudes, 
however, the sensor pointing at $\theta = 0^{\circ}$ significantly deviated from the
actual north direction. During the passages over the Sun's polar regions,
the sensor
always scanned through a plane tilted by about $30^{\circ}$ from the ecliptic
plane and all rotation angles lay close to the ecliptic plane 
(Figure~\ref{fig_orbitplot}).

The geometric detection probability for dust particles is defined by the
sensitivity of the detector for particles impinging from different directions
in an isotropic flux of particles. Directions are defined by the impact
angle, $\phi$, with respect to the sensor axis. The sensitive area as a 
function of $\phi$ is basically a 
cosine function modified by the shielding of the detector side 
wall \citep{gruen1992a}. The maximum area of $\rm 0.1\,m^2$ is found for
$\phi = 0^{\circ}$, and the sensor field-of-view is a cone with 
$70^{\circ}$ half angle.  
The solid angle covered by the detector is 1.45\,sr.
In an isotropic flux, 50\% of the particles hit the detector
at $\phi < 32^{\circ}$, while the impact direction of a single particle is
only known to be somewhere within the
$\mathrm{140}^{\circ}$ wide field-of-view. The average of all the rotation 
angle arrival directions of dust particles belonging to a stream 
is known to much higher accuracy than is the impact directon of 
a single particle.

Because of the mounting of the dust detector almost perpendicular to the
spacecraft spin axis,
the effective sensor area for dust impacts depends on the angle between
the impact direction and the spin axis. The maximum
sensitive area of the detector averaged over one spacecraft
revolution is $\rm 0.02\,\mathrm{m}^2$ \citep{gruen1992b}. 

Laboratory experiments showed that the sensor side wall was as sensitive to dust 
impacts as the target itself \citep{willis2005}, and candidates for wall impactors 
were indeed identified in the Ulysses interstellar dust data  \citep{altobelli2004a}.
While relaxing directional constraints, the wall impactors are not likely to  change our 
conclusions on grain sizes. The charge $Q_I$ measured 
on the ion collector of the dust instrument did not significantly differ between 
impacts onto the target and the sensor side wall. 

\subsection{Dust Impact Identification}

\label{sec_impact_identification}

The Ulysses sensor implements a highly reliable coincidence scheme of impact identification. 
Electrical signals in three independent channels arriving from a single dust impact are measured 
within less than 1~ms by two different methods (two charge-sensitive amplifiers and one multiplier). 
The amplitude ratios, the rise times, and the coincidence times are checked with reference 
to values that were obtained in calibration experiments, and true impacts are separated 
from noise events, the latter mostly trigger only a single channel.

Each measured signal (noise event or dust impact) was classified according to the strength of its ion 
charge signal ($Q_I$) into one of six amplitude ranges. Each amplitude range corresponds to roughly 
one decade in electronic charge $Q_I$. In addition, each event was classified into one of four event 
classes. The event classification scheme, defining the criteria to be satisfied for each class, is given 
in \citet{baguhl1993a} and \citet{krueger1999b}. This classification scheme was used for a 
reliable separation of noise events from true dust impacts. Real dust impacts had at least two
charge measurements plus additional coincidence criteria that had to be fulfilled 
({\em cf.} Table~\ref{tab_parameters}).

Four classes, together with six amplitude ranges, represent 24 separate categories. Each of these categories 
had its own 8 bit counter. Each signal registered by the dust instrument (noise event or dust impact) was 
counted with one of these counters even if the complete data set of the measured impact parameters 
(charges, rise times, coincidences, impact direction, etc.) was not transmitted to Earth. During periods 
of very high dust impact rates a small number of data sets was lost due to the 
limited data transmission rate of Ulysses (see also
Section~\ref{sec_identification}). The counter values, however, were always transmitted so that 
impact rates could be reconstructed.

\subsection{Dust Instrument Operation}

\label{sec_operation}

The Ulysses dust detector was operated almost without interruption from 
launch in 1990 until 2001. Due to decreasing power generation of the radioisotope batteries 
(RTGs), however, the available electrical power on board the spacecraft became an issue  
in 2001. Some instruments on board had to be switched off temporarily, and 
a cycling instrument operation scheme had to be 
implemented: one or more of the scientific instruments had to be switched off
at a time. As a consequence, the dust instrument was 
switched off repeatedly \citep{gruen1995c,krueger1999b,krueger2001b,krueger2006b,krueger2010b}.  
After 30 November 2007 the dust instrument remained switched off 
permanently even though the Ulysses spacecraft operation continued until 30 June 2009.

Degradation of
the dust instrument electronics, in particular the channeltron, was 
continuously monitored during the mission. 
We observed a channeltron degradation after approximately ten years of operation 
which was counterbalanced by an increase of the channeltron 
high voltage. We did not identify any other indications for instrument ageing in the 
Ulysses dust data.  
The smooth and rather undegraded behaviour of the Ulysses dust instrument is in contrast 
to the twin instrument on board Galileo: The electronics of the Galileo instrument suffered 
severe degradation due to the harsh radiation environment in Jupiter's magnetosphere
\citep{krueger2005a}, nevertheless, the coincidence scheme provided reliable impact 
identification even then.


\section{Identification of Interstellar Dust Impactors}

\label{sec_identification}

During the entire Ulysses mission the full data sets of 6719 dust impacts 
(containing impact time, impact charges, charge rise times, impact direction etc.
for each dust impact, {\em cf.} Table~\ref{tab_parameters}) were successfully transmitted 
to Earth \citep{krueger2010b}. 
During most time periods when the dust detector was operated, the impact rates were
sufficiently low so that the data sets of all recorded impacts could be transmitted 
to Earth. Only around the Jupiter flybys in 1992 and 2004 were very high impact rates 
 recorded so that the data sets of a large fraction of the detected impacts could not be 
 transmitted to Earth during
short intervals. All impacts, however, were always counted with the particle counters of
the dust instrument
\citep[{\em c.f.}, e.g.,][for details]{krueger2006b}. During the time intervals of interstellar
dust measurements considered in 
this paper the data sets of all recorded impacts were transmitted. 

An analysis of the dynamical properties (flux, impact direction) of the interstellar dust grains
detected during the entire Ulysses mission is given by \citet{strub2015}, and theoretical
predictions for interstellar dust flux, flow direction and mass distribution for Ulysses are studied
by \citet{sterken2015}. In the
present work we analyse the mass distribution of the interstellar grains as derived from the 
entire mission. To this end, we first 
have to identify the interstellar impactors in the Ulysses data set. 

\subsection{Dust Grain Dynamics}

\label{sec_dynamics}

The dynamics of interstellar dust grains in the solar system is dominated by three major forces:
solar gravity, solar radiation, and the Lorentz force. Here we briefly discuss the most important aspects
for particle motion in the solar system, a more comprehensive discussion is given
in the accompanying paper by \citet{sterken2015}.

Micron-sized and sub-micron sized dust particles are susceptive to a pressure exerted by the solar 
radiation field. Given that the solar radiation expands with the inverse square of the heliocentric 
distance, $r$, the radiation pressure force $F_{\rm RP}$ follows the same distance dependence as solar 
gravity $F_{\rm grav}$ (i.e. $r^{-2}$). Hence the ratio $\beta = F_{\rm RP}/F_{\rm grav}$ is constant 
for a given particle and depends on particle size, optical properties, morphology, etc. The radiation
pressure is strongly size-dependent, with a broad maximum for grain sizes approximately 
comparable to the wavelength of the incident radiation, i.e. for sub-micron grains. For strongly
absorbing materials, the $\beta$ ratio can be larger than one. 

Interstellar particles with $\beta > 1$ are deflected by the solar radiation, leading to an avoidance 
cone close to the Sun. For $\beta = 1$ 
radiation pressure and gravity cancel out and the particles move on straight ''undisturbed'' trajectories.
Particles with $\beta < 1$ are concentrated downstream from the Sun. 

Dust particles in interplanetary space usually carry an electric charge due to photoionization by the
solar radiation field, making them susceptible to the Lorentz force exerted by their motion through
the interplanetary
magnetic field. The field strength, orientation w.r.t. the particle motion and the grains' 
charge-to-mass ratio, $Q/m$, determine the strength of the Lorentz force. 
The surface charge of a spherical grain increases linearly with the grain radius, $a$,
while the mass has an $a^3$ dependence. Hence, the relative strength of the Lorentz force strongly 
increases for smaller particles. For a more detailed discussion the reader is referred to \cite{sterken2015}.

For the conditions in interplanetary space 
the Lorentz force becomes the dominating
force for particles smaller than approximately $a \lesssim \mathrm{0.1\,\mu m}$ 
\citep[][his Fig.~3.5; note that this value strongly depends on heliocentric distance]{landgraf1998a}. 
Particles larger
than approximately $1\,\mathrm{\mu m}$ have a low charge-to-mass ratio and low $\beta$, and their
dynamics is dominated by gravity. In the intermediate size range, radiation pressure makes a significant contribution, and it may even become the dominant force between 0.1 and $\mathrm{0.4\,\mu m}$ \citep{kimura1999}. 

The Lorentz force depends 
on the 22-year solar (magnetic) cycle, leading to a focussing and defocussing configuration for
interstellar dust.  
Particles with sufficiently 
high $Q/m$ are likely not able to penetrate the heliopause \citep[][note that these authors studied only 
the defocusing phase of the solar cycle]{linde2000}. 

We ignore the Poynting-Robertson drag force which is due to an abberation effect exerted on dust 
grains by the solar radiation field. It causes approximately micrometer-sized particles initially 
orbiting the Sun at 1\,AU distance to spiral into the Sun on timescales of $10^3$ to $\mathrm{10^4\,years}$. 
The interstellar grains traverse the solar system
within 20 to 50 years and they spend a much shorter time close to the Sun where the Poynting-Robertson
drag is strongest. The resulting grain deflection is very small and thus negligible in our case.

We ignore any rotation of the dust grains that might lead to rotational bursting of the grains.
Rotational bursting which might have an
affect on the dust mass distribution by creating an excess of small grains, is not expected for  
interstellar grains in the heliosphere \citep{misconi1993,draine2011}. Similarly, the Yarkovsky effect 
can be ignored for the 
interstellar dust grains \citep{gustafson1994}. 

Furthermore, one might expect a contribution from the
rotational energy of the grains to the energy released during impact onto the detector target. The rotational 
energy of micrometer and submicrometer sized grains is 10 to 15 orders of
magnitude smaller than the kinetic energy of the grains. Hence, the rotational energy 
can be completely ignored for the calibration of our impact measurements. We also ignore
any other mechanisms of grain destruction \citep[see][their Sect. 4.4]{frisch1999a}.  

\subsection{Grain selection criteria}

\label{sec_selection}

The Ulysses dust data contains impacts by interplanetary as well as interstellar grains. Therefore, 
we have to find selection criteria that allow us to define data sets with a negligible number of
impacts from sources other than interstellar. 
For the identification
of the interstellar impactors we have 
adopted the same selection criteria as \citet{landgraf1998a} and \citet{frisch1999a} 
and applied them to the entire Ulysses mission. These criteria were based on the following observations
that we are reviewing here:  
\begin{itemize}
\item[1)] After its Jupiter flyby in February 1992, Ulysses observed a relatively 
constant flux of dust particles above and below the ecliptic plane. The approach direction
of these grains was opposite to the direction of interplanetary dust during most of the time, 
except around Ulysses'  perihelion. Hence, they appeared to be 
in retrograde motion about the Sun (Figure~\ref{fig_orbitplot}). If we assume that these grains enter the solar system
from close to the upstream direction of the interstellar helium gas as observed by Ulysses,
their impact direction is compatible with an origin from outside the solar 
system. 
\item[2)] Applying the mass and speed calibration of the dust instrument, most particles had
impact speeds in excess of the solar system escape speed, also pointing to an origin from 
outside the solar system. 
\item[3)] The flux of the interstellar particles was independent of ecliptic latitude 
\citep{landgraf2003,krueger2007b} in 
contrast to interplanetary dust that is strongly concentrated toward the ecliptic plane
and the inner solar system. Dust emanating from the jovian system is concentrated in the
vicinity of Jupiter \citep{gruen1993a,krueger2006c}.
\end{itemize}


From the above observations we derive the following identification criteria for  interstellar
grains: From observation \#1 we select every impact that was measured when the interstellar 
helium flow direction was within the $\pm 70^{\circ}$ field-of-view of the dust detector. We add 
for a $20^{\circ}$ margin because 
the sensor side wall turned out to be as sensitive to dust impacts as the target itself 
(Section~\ref{sec_angular_sensitivity}). 
When Ulysses crossed the ecliptic plane at a heliocentric 
distance of about 1.3~AU, the impact directions of interstellar and prograde interplanetary grains
were not as clearly separated as it was the case during the rest of the Ulysses orbit. We therefore
exlude all impacts around perihelion when Ulysses was between $-60^{\circ}$ and $+60^{\circ}$ ecliptic 
latitude. 

Over the poles of the Sun, Ulysses detected very small particles which were interpreted as fragments of 
interplanetary grains ejected from  the inner solar system by electromagnetic effects \citep{hamilton1996a}
and solar radiation pressure \citep{wehry1999,wehry2004}. 
In order to remove these particles from the data set the measured amplitude of the ion charge 
signal, $Q_I$,  had to be more than one order of magnitude above the detection threshold of the dust 
instrument. Therefore, over the poles of the Sun, at ecliptic latitudes $|b| \geq 60^{\circ}$,
we ignore impacts with impact charge amplitudes $ Q_I \leq 10^{-13}\,\mathrm{C}$. 

Around the Jupiter flybys in 1992 and 2004 Ulysses detected collimated streams of dust particles 
originating from within Jupiter's magnetosphere \citep{gruen1993a}. In order to avoid
contamination of the interstellar dust data set, the measurements during the periods of 
identified jovian dust streams were ignored entirely. The times when the dust streams occurred were
given by \citet{baguhl1993a} and \citet{krueger2006c} and are adopted here. 

\begin{figure}
   \centering
\parbox{15cm}{
\vspace{-5cm}
\includegraphics[width=0.8\textwidth]{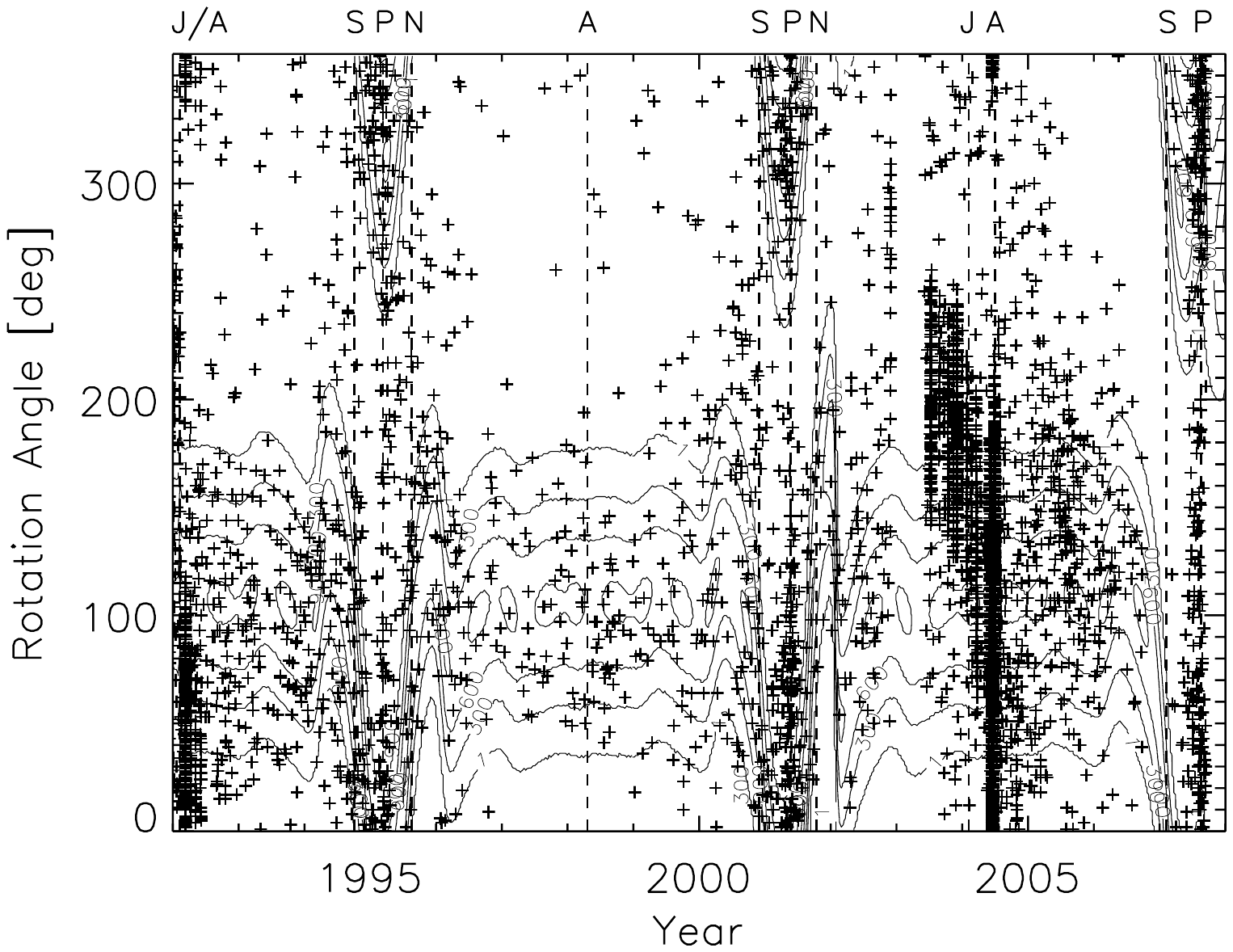}
\vspace{-11cm}
}
\parbox{15cm}{
\hspace{1mm}
\vspace{-5cm}
\includegraphics[width=0.8\textwidth]{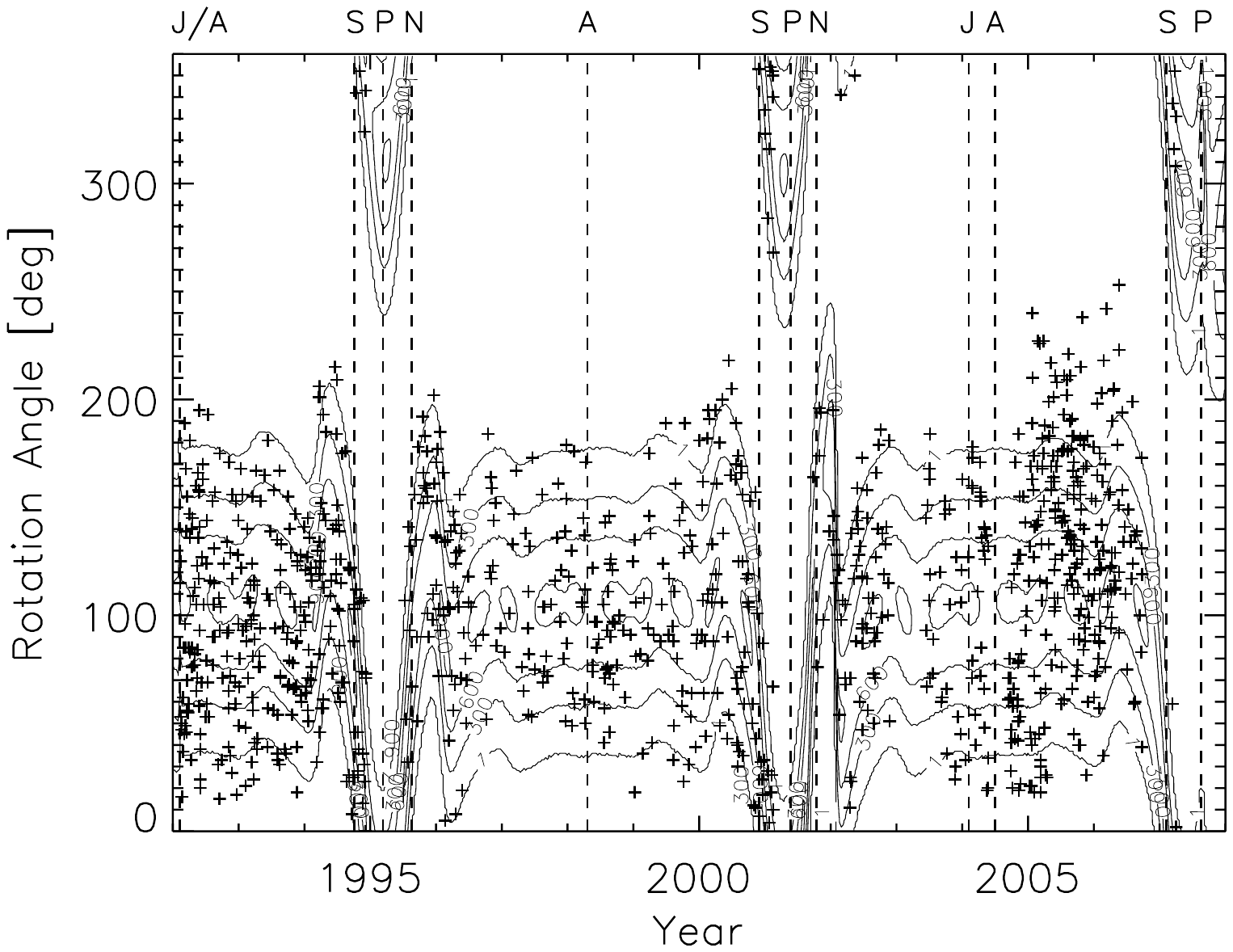}
}
\hspace{1mm}
\vspace{-0.5cm}
\caption[]{Impact direction (rotation angle) vs. time for all dust impacts detected between Jupiter flyby in February 1992 and the end of dust instrument operation in November 2007 (top) and for the
identified interstellar dust impactors (bottom). Each cross indicates an individual dust particle impact. Contour lines show the effective sensor area for
dust particles approaching from the upstream direction of interstellar helium
\citep{mccomas2012}. Vertical dashed lines and labels at the top indicate Ulysses' Jupiter flybys (J), perihelion passages (P), aphelion passages (A), south polar passes (S) and north polar passes of Ulysses (N). 
}
\label{fig_rotplot}
\end{figure}

Finally, a shift in the approach direction of the interstellar grains by about $40^{\circ}$ 
was recognised in 2005 and 2006 \citep{krueger2007b,strub2011,strub2015}. Therefore in 2005/2006 the
nominal band of rotation angles of $\pm 90^{\circ}$ within the interstellar helium flow direction was 
expanded  towards larger rotation angles by $40^{\circ}$ to take this shift into account 
({\em c.f.} Figure~\ref{fig_rotplot}). 

The selection criteria for the identification of interstellar grains are listed 
in Table~\ref{tab_criteria}. Our criteria are different from those adopted by  
\citet{strub2015}. Since our aim is to derive the mass
distribution of the grains we have to use criteria that do not induce any bias in the mass 
distribution. Therefore, we did not constrain the measured impact charge, $Q_I$, except for short periods
over the poles of the Sun ({\em c.f.} Table~\ref{tab_criteria}). We used the observed impact 
direction of the interstellar grains by constraining the rotation angle. We do not expect to
introduce a bias in the mass distribution this way. On the other hand, \citet{strub2015} 
analyse the dynamical properties of the grains. For example, to avoid any bias in the measured
impact directions these authors did not constrain the rotation angle.

After removing potential impacts by interplanetary particles with the method described above,
we identified 987 interstellar grains in the Ulysses data \citep[compared to 526 interstellar grains
identified by][]{strub2015}. The Ulysses interstellar dust data 
is shown in Figure~\ref{fig_rotplot} (subsets of the Ulysses 
interstellar dust data for different size bins are shown 
in Figure~\ref{fig_rotplot2} in the Appendix).
This extends the number of detected interstellar grains 
 by more than a factor of three compared to earlier analyses \citep[305 Ulysses 
impacts between Jupiter flyby in February 1992 and March 1996;][]{landgraf1998a,frisch1999a}. 

The earlier works also
considered 309 interstellar dust impacts measured with Galileo. Here we use only the Ulysses data.
Inclusion
of the Galileo data would extend the entire data set by only approximately 1/3 and, hence, not seriously 
increase the statistical significance of our results. On the other hand, \citet{landgraf2000a} 
concluded that the Galileo data set is likely contaminated with impacts by interplanetary grains. 
Galileo  measured only in the ecliptic plane where a stronger contribution by interplanetary 
impactors has to be expected, while Ulysses measured out of the ecliptic plane most of the time. 


\subsection{Dust impact speed}

The grain impact speed derived from the instrument calibration can be considered as an 
independent consistency check of our grain selection criteria. Out of the data set of 987
interstellar grains identified by the selection criteria described in 
Section~\ref{sec_selection} we selected only those grains with a reliable measurement
of the charge rise times, $t_I$ and $t_E$, and, hence, a reliable determination
of the grain impact speed 
\citep[i.e. velocity error factor $\mathrm{VEF < 6;}$][]{krueger2010b}.
This results in a data set of 943 particles. The average impact speed of these grains
is $\mathrm{24 \pm 12 \,km\,s^{-1}}$, confirming earlier results by \citet{kimura2003a}. 
Even though this value is very close to the measured speed of the interstellar gas 
of 23 to $\mathrm{26\,km\,s^{-1}}$, it should not be taken as a discriminator for either of
the two values due to the factor or two uncertainty in the measurement of the interstellar dust speed. 

From modelling the particle dynamics \citep[e.g.][]{sterken2015}
the largest grains are expected to have the highest impact speed due to gravitational acceleration, 
the mid-sized particles with sizes close to the maximum of the $\beta$ curve have smaller
impact speeds, and the smallest particles have variable speeds due to the Lorentz force. 
The measured average impact speed is in agreement with our hypothesis that the interstellar 
dust flow is generally coupled with the gas, and 
that the majority of the grains in our selected data set are indeed of interstellar origin.

\subsection{Determining the Dust Mass Distribution}

\label{sec_massdist}

The most straightforward determination of the grain mass is based on the 
laboratory calibration of the dust instrument and relies on 
the grain impact speed as derived from the measured rise time of the charge signal 
({\em c.f.} Section~\ref{sec_calibration}). Equation~\ref{eq_charge} shows that the
mass obtained from the impact charge measurement
has a strong dependence on the impact speed, with a power law index of approximately $3.5$.
Given that the speed calibration has a factor of two uncertainty, 
this yields a factor of ten uncertainty in the derived mass. 

A more accurate mass can be derived if the grain impact speed is known by other 
means. Such a technique was successfully applied earlier to Ulysses interstellar
dust measurements by \citet{landgraf2000a} and to Galileo dust measurements 
of grains ejected from the Galilean moons \citep{krueger2000a,krueger2003b}.
In the present work we apply two similar approaches to determine the  
impact speed of the interstellar dust grains.
Both take into account the change in velocity of an interstellar grain in the heliosphere 
which can -- in principle -- easily be determined from 
the acceleration due to solar gravity and radiation pressure. We neglect the Lorentz force exerted 
on the grains by interaction with the solar wind magnetic field, which is a good approximation for grains
more massive than approximately $10^{-16}\mathrm{\,kg}$. The Larmor radii for such 
particles are on the order of 500~AU in the region traversed by Ulysses, increasing with
distance from the Sun \citep{gruen1994a}. They are much larger than the length of their interaction
with the solar wind. 

The relative strength of radiation pressure is
expressed as the ratio, $\beta$,
between the radiation pressure force and the gravitational force 
(Section~\ref{sec_dynamics}). For 
sub-micrometer grains radiation pressure can be of the same order   
($\beta \approx 1$) or even larger than gravity ($\beta > 1$). We therefore consider two
simple cases, following the strategy  applied by \citet{landgraf2000a}: 

\begin{itemize}
\item[Model 1)] The radiation pressure force and gravity acting on a dust grain 
have exactly the same strength but opposite 
directions ($\beta=1$, fixed). Therefore, the interstellar grains move through the solar system on 
straight lines.  
Their velocity and flow direction remain unchanged. In this case, the impact velocity is given 
by the difference between the grain velocity at infinity and the spacecraft velocity.
\item[Model 2)] The ratio $\beta$ depends on the grain size. In this case, the grain velocity is affected by radiation 
pressure and gravity. We calculate $\beta$ and the grain velocity for each grain 
individually. We take $\beta$ from \citet{kimura1999} as a function of grain radius, $a$, for
compact spherical grains
made of astronomical silicates, having a bulk density of $\rho = 3.3 \times 10^3\mathrm{\,kg\,m^{-3}}$. The 
grain radius, however, is not measured independently by the dust instrument. We therefore have to 
derive the radius from the grain mass, $m$, obtained from the impact charge measurement. The radius
of a spherical grain is given by $a=(3 m/( 4\pi \rho))^\frac{1}{3}$. Using the grain radius, we can determine 
the dust velocity in the heliocentric frame
and hence the impact velocity onto the sensor target. Then, an improved grain mass can be 
calculated by inversion of Equation~\ref{eq_charge}. From this mass we determine a new grain radius which gives
us a new $\beta$, and so forth. This iterative process leads to a value of $\beta$ in a self-consistent way. 
\end{itemize}

A disadvantage of this second method is its dependence on the detailed properties of the dust 
grains which are not well known. We therefore apply both models and compare the results.
It turned out that $\beta = 1$ is a good approximation for the majority of the impacts detected with 
Ulysses and Galileo \citep{landgraf2000a}. For the biggest detected grains and for the smallest ones, 
however, this is not a good approximation. Here, the second model 
is expected to give better results.

For both models we assume an initial velocity of the grains outside the heliosphere of 
$23.2\,\mathrm{km\,s^{-1}}$ \citep{mccomas2012}. This value is about 10\% smaller than the 
value of $\mathrm{26\,km\,s^{-1}}$ that was earlier adopted by \citet{landgraf2000a}. 
Equation~\ref{eq_charge} shows
that the derived grain masses increase by about 50\%  due to this reduced impact speed.

We assume an upstream direction of the interstellar dust flow of $\mathrm{250^{\circ}}$ ecliptic
longitude and $\mathrm{8^{\circ}}$ latitude as was recently derived by \citet{strub2015}. 
The longitude is somewhat smaller than the value derived by \citet[][$\mathrm{259^{\circ}}$]{landgraf1998a}.
Given the large field-of-view of the dust detector, this is well within the measurement 
uncertainty.
For most of the time, except in 2005/06, this initial velocity vector is 
(1) compatible with the heliocentric speed and the direction of motion of the interstellar grains 
detected with Ulysses 
\citep{gruen1994a,baguhl1995b}, 
(2) close to the asymptotic velocity vector of the interstellar helium
flow detected by Ulysses and IBEX \citep[][Section~\ref{sec_introduction}]{witte2004a,mccomas2012}, and 
(3) close to the velocity of the Sun with respect to the local interstellar cloud \citep{lallement1992}. 
In 2005/06 we take a $\mathrm{40^{\circ}}$ shift in 
the grain impact direction into account \citep{strub2015}.

A recent analysis of neutral helium
measurements revealed a potential temporal variation of the inflow direction and speed of neutral helium 
over four decades \citep{frisch2013b}, which was later put into question by \citet{lallement2014} who found no evidence for such a variation.
For the measurement period of the Ulysses interstellar dust measurements 
this corresponds to a shift of $2.7^{\circ}$ over 16 years. Given the large field-of-view of the dust 
detector (Section~\ref{sec_angular_sensitivity}), this value is negligible for our analysis. 


\section{Results}

\label{sec_results}

\subsection{Dust Mass Distribution}

\begin{figure}
   \centering
   \vspace{-1cm}
   \parbox{15cm}{
\vspace{-3.5cm}
\includegraphics[width=0.52\textwidth]{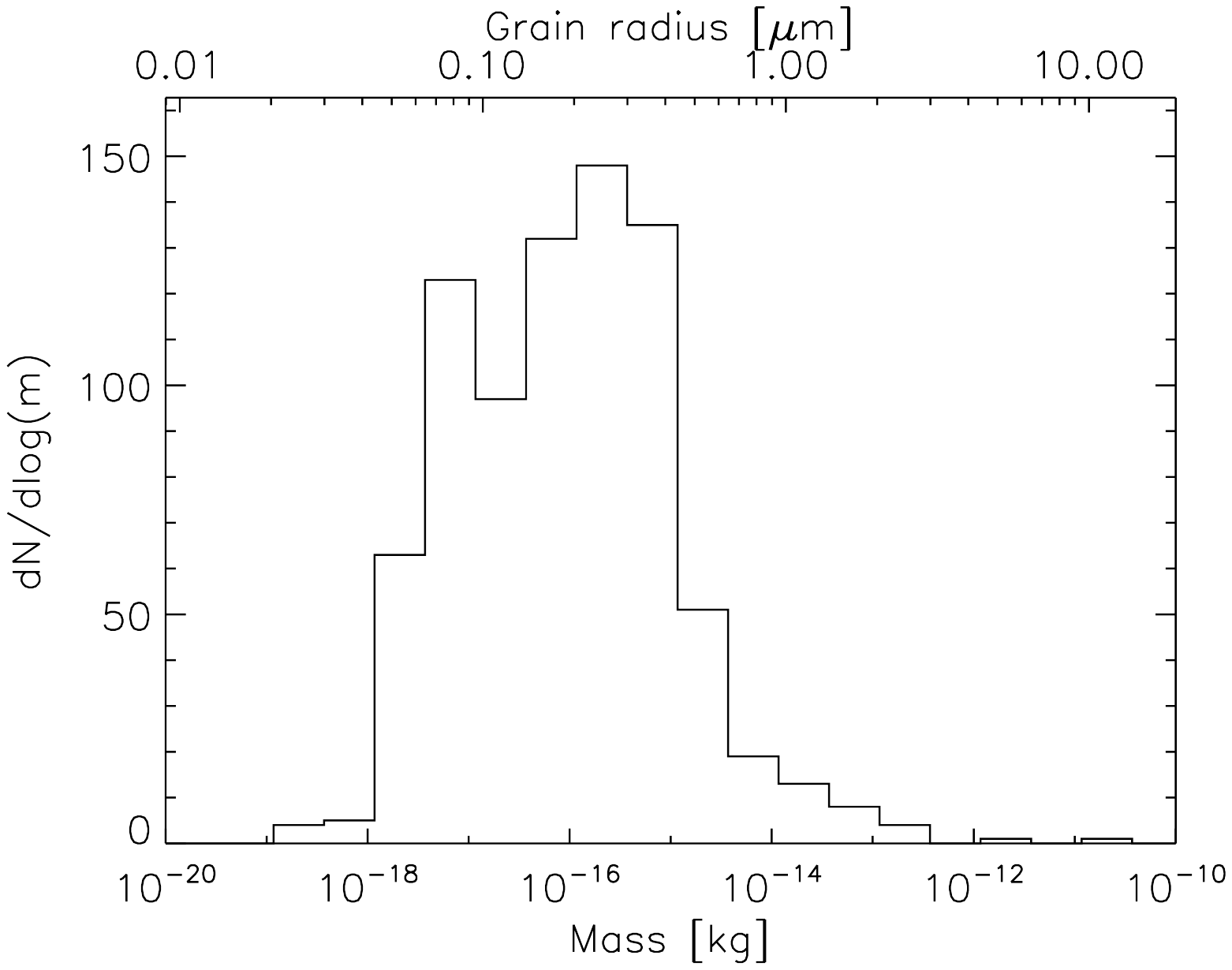}
\vspace{-4.5cm}
}
\parbox{15cm}{
\hspace{0.1mm}
\vspace{-1.5cm}
\includegraphics[width=0.5\textwidth]{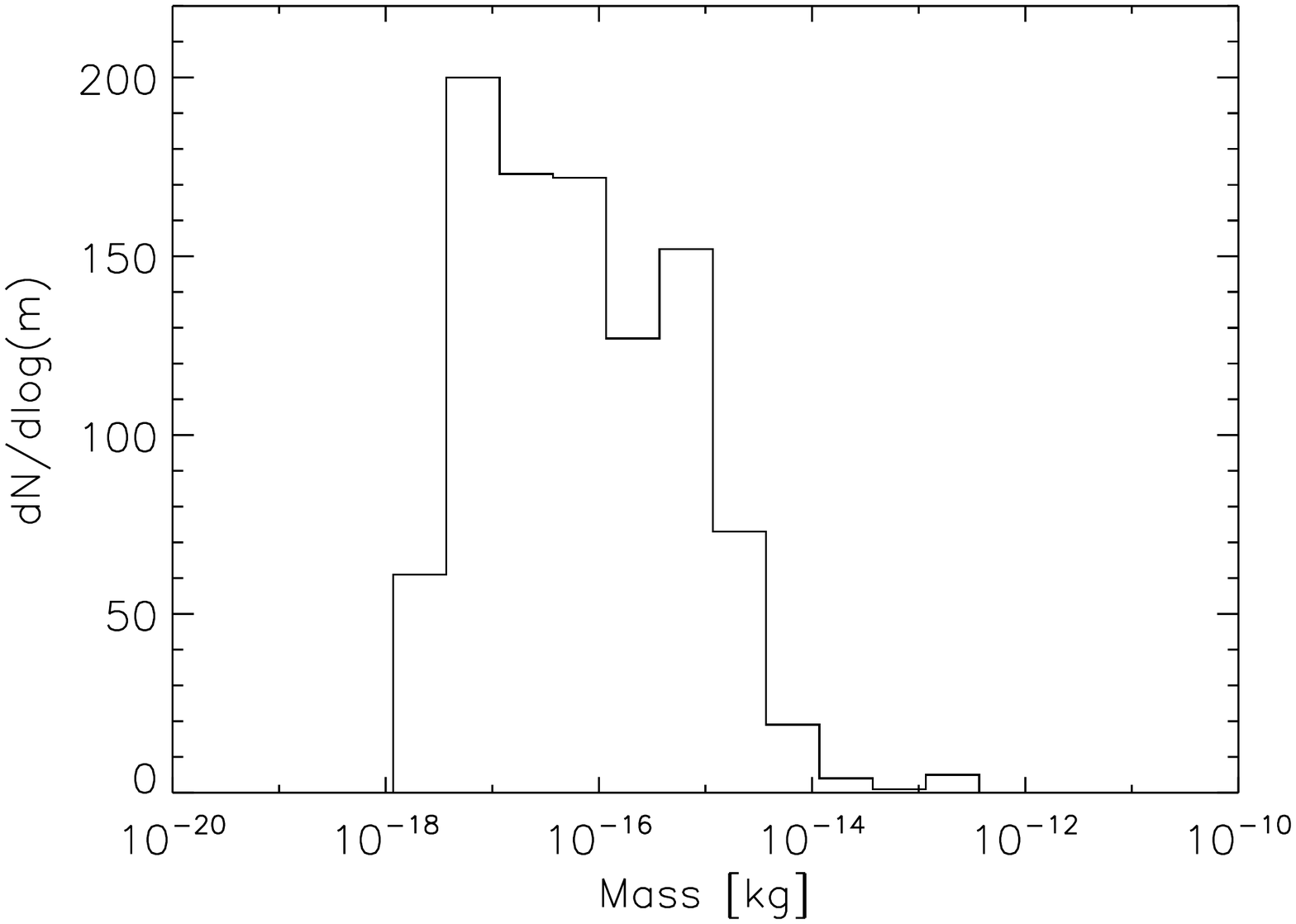}
}
\parbox{15cm}{
\hspace{0.1mm}
\includegraphics[width=0.505\textwidth]{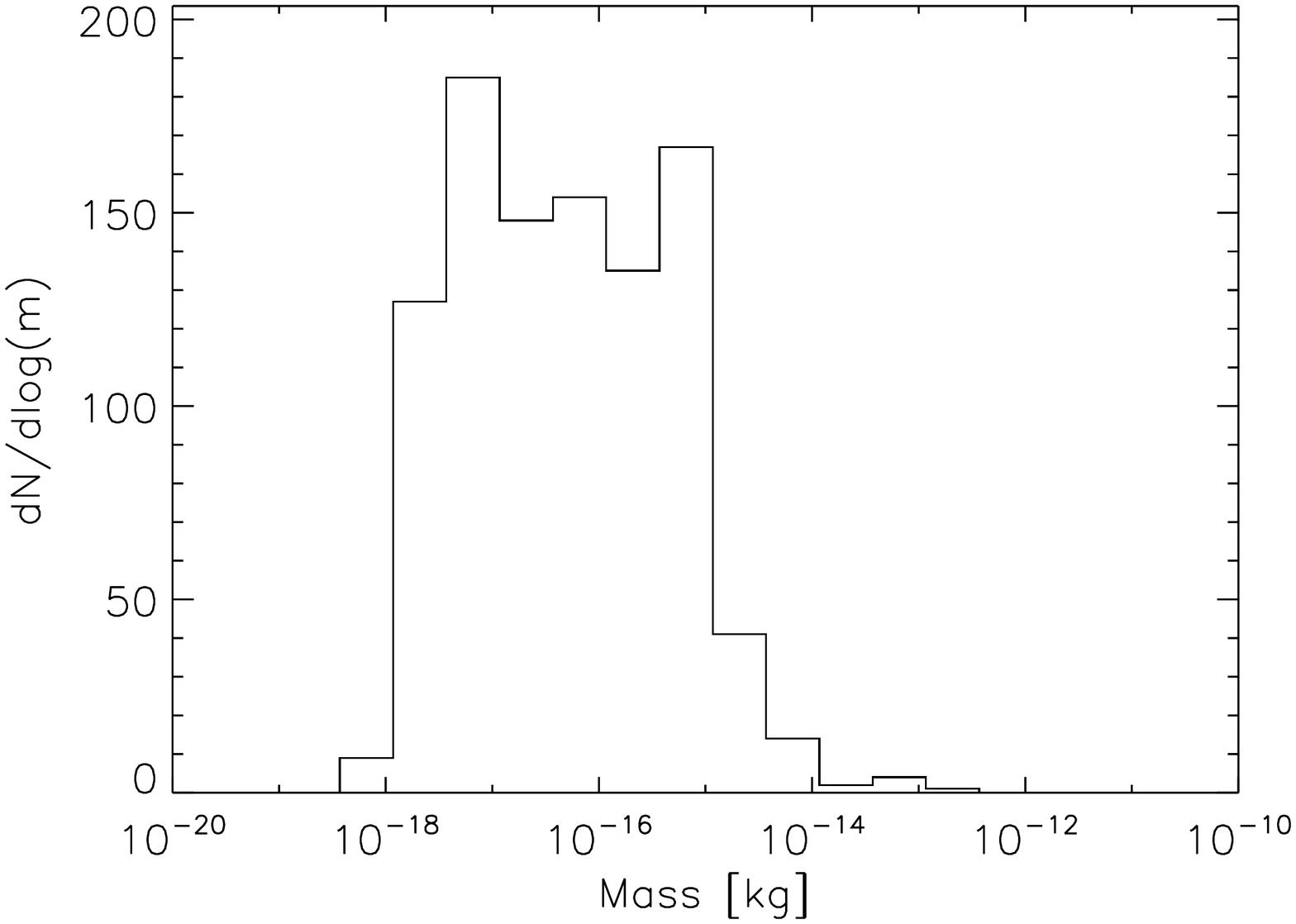}
}
\hspace{-1mm}
\vspace{-0.5cm}
\caption[]{Mass distribution of interstellar grains derived from the Ulysses measurements 
shown as number of particles per logarithmic 
mass interval for three different cases for the impact speed calculation. 
{\em Top panel:} Grain masses derived from the measured impact speeds.  
Only particles with impact speed $v>13\,\mathrm{km\,s^{-1}}$ were considered (804 particles).
{\em Middle panel:}  Masses derived from the $\beta = 1$ model, taking into account the spacecraft motion (model 1).
{\em Bottom panel:} Masses derived self-consistently (model 2) with accelerated ($\beta < 1$) and 
decelerated ($\beta > 1$) grains (987 particles for both models). The approximate grain size for spherical 
particles with density $\rho = 3.3 \times 10^3\,\mathrm{kg\,m^{-3}}$ is shown at the top for comparison.}
\label{fig_massdist}
\end{figure}

The resulting mass distributions for the three cases considered (calibrated impact speed, $\beta = 1$ (model 1),
and $\beta$ variable self-consistent (model 2)) are shown in Figure~\ref{fig_massdist}. They cover a mass range from 
approximately $10^{-18}\,\mathrm{kg}$ (which is the detection threshold for grains impacting with 
$20\,\mathrm{km\,s^{-1}}$) to $10^{-10}\,\mathrm{kg}$ with maxima at about 
$10^{-17}\,\mathrm{kg}$ to $10^{-16}\,\mathrm{kg}$. From modelling the extinction of starlight 
\citep{mathis1977,draine2009} it is expected that the number of grains per mass interval steeply rises 
towards smaller grain masses. This is not seen in the in-situ data, instead the mass distribution
shows a deficiency of small grains below approximately $10^{-16}\,\mathrm{kg}$ (top panel).
This deficiency is most likely due to the interaction of the grains with the  
interplanetary magnetic field \citep{gruen1994a}. The upper mass limit at approximately 
$10^{-11}\,\mathrm{kg}$ is determined by the size of the dust detector: Large grains are much less
abundant than small ones so that only very few large grains were detected.

Comparison of the top panel with the two lower panels 
in Figure~\ref{fig_massdist} shows that the proportion of particles below $10^{-16}\,\mathrm{kg}$
is increased and the fraction of particles above this limit is reduced when we derive the grain
masses from the $\beta=1$ model and the self-consistent model for the grain impact speed. A similar
result was also found  by \citet{landgraf2000a} from the analysis of the Galileo and the smaller Ulysses 
interstellar dust data sets available at the time. It was explained by being either due to a 
contamination by interplanetary impactors that might have lower impact speeds than the interstellar
grains or by recombination in the impact-generated plasma cloud in the detector. 

The mass distributions derived from the two impact speed models ($\beta=1$ and the self-consistent model) 
are very similar, except that the number of grains at the large mass end is even further
reduced in the self-consistent model ({\em c.f.} middle and bottom panels of  Figure~\ref{fig_massdist}),
again confirming earlier results by \citet{landgraf2000a}.  
 
We now consider the contributions of grains with different masses to the overall mass density of 
interstellar dust in the solar system. In Figure~\ref{fig_masshist} we show the mass distribution of 
interstellar grains as the differential mass density per unit volume 
\citep[987 particles; see also][]{frisch1999a}. The distribution derived from astronomical
observations \citep[][hereafter MRN]{mathis1977} for an interstellar hydrogen density of 
$\mathrm{0.22\,cm^{-3}}$ 
is shown for comparison. Particles with masses below approximately $10^{-16}\,\mathrm{kg}$ are
strongly depleted in the inner heliosphere  due to heliospheric
filtering, as compared to the interstellar medium. For instance, 
the density of grains with mass $10^{-17}\,\mathrm{kg}$ is reduced in
the inner heliosphere by about a factor of 90 below the MRN prediction while 
$10^{-18}\,\mathrm{kg}$ grains are deficient by three orders of magnitude. 
At the same time large (approximately $10^{-14}\,\mathrm{kg}$) grains are absent in
the MRN distribution, but are abundant in the inflowing interstellar dust. 
It is incompatible with both
interstellar elemental abundances and the observed extinction properties of the interstellar
dust population \citep{draine2009}. The Solar System may by chance be located near a
concentration of massive grains in the interstellar medium \citep[$\mathrm{\ll 1\,kpc}$;][]{gruen2000b}. 

The existence of  interstellar grains larger than approximately $10^{-16}\,\mathrm{kg}$ as derived 
from the Ulysses 
and Galileo  data was an important result from the earlier interstellar dust measurements. 
The largest contribution of the detected grains  to the optical cross-section is provided by 
grains in the range $10^{-16}\,\mathrm{kg}$ to $10^{-14}\,\mathrm{kg}$, while smaller grains below 
$10^{-16}\,\mathrm{kg}$ that are believed to dominate the extinction of starlight
do not contribute much to the mass density \citep{landgraf2000a}. Such small grains are 
significantly depleted in the Ulysses data due
to interaction with the interplanetary magnetic field and the heliospheric boundary 
during certain time intervals \citep{slavin2008,slavin2010}.
On the other hand, the large grains above $10^{-16}\,\mathrm{kg}$  
provide a significant contribution to the total mass of dust in the interstellar medium,
given their large masses and relative abundance. 

The total mass density of interstellar grains as derived from the Ulysses in-situ data
can be obtained by integrating over the differential distribution shown in Figure~\ref{fig_masshist}.
This yields a total mass density of $(2.1 \pm 0.6)\times 10^{-24}\,\mathrm{kg\,m^{-3}}$ which 
is a factor of three smaller than the value derived  by \citet{landgraf2000a}. This value
is dominated by the largest particles detected \citep[see][their Fig.~7c]{landgraf2000a}. 
The reduced dust density reflects a smaller proportion of the biggest grains detected after 2000,
assuming there are no small-scale variations in the dust density in the ambient
interstellar medium close to our solar system.
Temporal variations in the flux of these large grains
are not likely, as they are only marginally affected by the 
time-variable interplanetary magnetic field. On the other hand, the dust density varies 
spatially as the large grains are focussed in the downstream direction behind the Sun. When
Ulysses moved towards the Sun, a dust density increase by a factor of 1 to 1.5 and a relative 
increase in interstellar flux by a factor of 2 to 2.5 with respect to the undisturbed 
incoming density and  flux are expected from simulations \citep{sterken2015}. 
However, these regions around Ulysses' perihelion are ignored in the data selection
so that this has a minor effect on the derived mass distribution.

\subsection{Gas-to-dust mass ratio in the local interstellar cloud}

From the total mass density derived from the in-situ measurements we can calculate the gas-to-dust
mass ratio in the local interstellar cloud surrounding our solar system. It gives us information 
about the refractory elements
in our local interstellar environment. We adopt a recently determined total hydrogen density of 
$n_{\mathrm{H}}=0.247\mathrm{\,cm^{-3}}$ 
\citep[i.e. neutral hydrogen density $n_{\mathrm{HI}}=0.192\, \mathrm{cm^{-3}}$ and proton 
density $n_{\mathrm{p}}=0.0554\,\mathrm{cm^{-3}}$;][their model 26]{slavin2008}, and
a neutral helium density of $n_{\mathrm{He}} = 0.015\, \mathrm{cm^{-3}}$ 
 \citep{moebius2004}. 
Using the total dust mass density derived 
from the interstellar grains detected with Ulysses (Section~\ref{sec_massdist}), 
we find a gas-to-dust mass ratio in the local interstellar cloud of $R_{g/d} = 193^{+85}_{-57}$. 
This value is somewhat higher than the dust density derived from earlier investigations  
\citep[$R_{g/d} \sim 94 - 127$;][]{frisch1999a,landgraf2000a,kimura2003a,altobelli2004a}. 
It should be mentioned that there is some uncertainty in the total hydrogen density. 
For example, \citet{heerikhuisen2011}, from heliosphere models, find a somewhat lower value of
$n_{\mathrm{H}}=0.21 - 0.23\mathrm{\,cm^{-3}}$. In our analysis, a value 
of $n_{\mathrm{H}}=0.22\mathrm{\,cm^{-3}}$ results in $R_{g/d} = 172$.

Gas-to-dust mass ratios calculated from more recent models with improved solar abundances are in the 
range $R_{g/d} \sim 149 - 217$  \citep{slavin2008}. Thus, our
present analysis is in good agreement with the results obtained from 
astronomical observations. 

\subsection{Interstellar dust flux}

In Figure~\ref{fig_flux} we show the cumulative mass flux as derived from the Ulysses interstellar
dust measurements. Here we show only the self-consistent model for the speed calibration (model 2). For
a discussion of the two other alternatives for calibrating the grain masses the reader is
referred to \citet{landgraf2000a}. The dust flux distribution extends 
to somewhat larger particles as compared to the earlier analysis by \citet{landgraf2000a} 
for
two reasons: (1) The reduced impact speed in our present analysis leads to 
larger grain masses, and (2) the dust data set contains about a factor of three more 
particles so that the dust detector had a higher chance to catch larger particles. 
The flux of $10^{-13}\,\mathrm{kg}$ particles
is on the order of $\mathrm{10^{-7}\,m^{-2}\,s^{-1}}$.


\section{Discussion}

\label{sec_discussion}

The STARDUST mission recently returned samples of contemporary interstellar grains to Earth. 
Preliminary analysis of a few of these grains extracted from
the  interstellar collector indicates that their bulk density  is
rather low \citep{westphal2014b}. Also \citet{sterken2015} conclude from the simulations 
in the context of Ulysses observations on low density interstellar dust. The bulk density 
affects the charge-to-mass
ratio for a given size and, hence, the grain interaction with the interplanetary 
magnetic field.

In our analysis we assumed the grains to be 
compact, spherical and composed of astronomical silicates with density 
$\rho = 3.3 \times 10^3\,\mathrm{kg\,m^{-3}}$ \citep{kimura1999}. We did not
consider porous grains for three reasons: (1) the bulk density is not yet 
well established from the analysis of 
the STARDUST samples; (2) the laboratory calibration of the Ulysses dust detector was 
performed solely with compact grains. Only recently are there attempts to calibrate the 
dust detector with low-density grains with the Heidelberg Dust Accelerator \citep{sterken2013b}; 
(3) Finally,
the $\beta$ curves for porous particles are presently  under review (Hiroshi Kimura, priv. 
comm.). Once these prerequisites are fulfilled, it will be possible to do the next major step 
in deriving a more consistent calibration of the interstellar grain mass distribution, 
matching also the STARDUST results. We estimate that if the interstellar grains are of low density indeed, their masses would be typically overestimated by one order of magnitude in the Ulysses data \citep{sterken2012b,sterken2015}. This would increase the gas-to-dust ratio 
calculated in this paper. On the other hand, many big particles, the flux of which peaks around perihelion 
\citep{sterken2015}, were left out of the selection (Section~\ref{sec_identification}). This could reduce 
the gas-to-dust mass ratio. It is not clear at this stage which effect is bigger, and this needs 
further investigations.

In addition to the well recognized silicate component of interstellar dust, astronomical observations 
also indicate the existence of carbon grains in interstellar space \citep{kimura2003b,draine2011}. Carbon 
has a higher albedo (i.e. higher $\beta$) than silicates and is thus more susceptible to radiation pressure. 
In order to test the influence of a significant carbon component in 
our Ulysses detections we assumed that all 
detected grains are composed of carbon, and we used 
the $\beta$ curves for compact carbon from \citet[][their Fig.~1]{kimura1999}, instead of
the silicate data. With this assumption we recalculated the grain masses with our self-consistent 
model with variable $\beta$. 
This leads to a reduction in 
the gas-to-dust mass ratio by about 20\%. It should be noted, however, that this is a very 
simple approach which neither takes into account an influence of the grain composition on the calibration 
of the impact measurements, nor porosity of the grains. 
Furthermore, we do not know the abundance of carbon grains in the
interstellar dust flow yet. The existence of the ${\mathrm 9.7\,\mu m}$ and ${\mathrm 18\,\mu m}$
infrared features observed in interstellar clouds indicates that silicate grains 
are abundant in interstellar space which is also consistent with the
STARDUST results \citep{westphal2014b}.

Similarly, the entry speed of the interstellar helium into the heliosphere was
under debate \citep{lallement2014,wood2015,mccomas2015}.  
 Values of $\mathrm{23.2\,km\,s^{-1}}$ and 
$\mathrm{26\,km\,s^{-1}}$, respectively, were considered. 
Our model with variable $\beta$ and an 
entry speed of the interstellar grains set to this latter value with all parameters unchanged, 
yields a gas-to-dust mass ratio  about 20\% higher than derived for the lower 
entry speed. 

Modelling of the interaction of the small interstellar grains with the solar wind magnetic 
field suggests that the mass distribution changes with time 
\citep{landgraf1999a,landgraf2003,sterken2013a}. 
Small grains are depleted between mid-1996 and 1999 because of the 
defocussing configuration of the solar wind magnetic field. The analysis of the Ulysses 
data suggests such a depletion of the interstellar grains in this time interval.
On the other hand, a concentration of big grains is expected in the downstream direction of the 
interstellar dust flow behind the Sun. Thus, the measured flux of big grains should have increased 
around Ulysses' perihelion passage when the spacecraft was close to this region.
We have ignored this time interval in our analysis because interstellar grains cannot be clearly 
separated from interplanetary impactors in this  period. 

We did not consider temporal changes in our analysis of the grain size distribution. Changes in
the slope of the mass distribution are
discussed in an accompanying paper by \citet{strub2015} which revealed 
temporal and grain-size dependent variations of the measured dust flux and impact direction. 
Simulations of the dust size and mass 
distributions for so-called adapted astronomical silicates show some features similar to 
the observed dust distribution \citep{sterken2015}.

Dust measurements between 0.3 
and 3~AU in the ecliptic plane exist also from Helios, Galileo 
and Cassini. These data show evidence for distance-dependent 
alteration of the interstellar dust stream caused by solar radiation 
pressure, gravitational focussing by the Sun and electromagnetic interaction 
of the grains with the time-varying interplanetary magnetic field 
\citep{altobelli2003,altobelli2005a}. 

The gas-to-dust mass ratio derived from 
our analysis is dominated by the largest grains detected. The largest grains, however, are 
not seriously affected by radiation pressure and electromagnetic forces. The neglect of potentially
big interstellar impactors in the inner solar system may lead to an overestimation of the gas-to-dust mass 
ratio $R_{g/d}$. We will address this aspect in detailed simulations of the grain dynamics 
\citep{sterken2015}. 



\section{Conclusions}

\label{sec_conclusions}

We analysed the mass distribution of interstellar dust grains entering the heliosphere from 
16 years of Ulysses in-situ dust measurements obtained 
between February 1992 and November 2007. Our analysis extends the time period  sampling the
interstellar dust size distribution in the heliosphere by more than a factor of two compared to 
previous investigations by 
\citet{landgraf2000a}.
A total number of 987 interstellar dust impacts was identified in the Ulysses dust data, thus 
extending the total interstellar dust data set by a factor of three compared to earlier 
analyses. 

We used a very similar technique as \citet{landgraf2000a}, however, with updated 
properties of the interstellar medium: 
interstellar dust  speed outside the heliosphere of $\mathrm{23.2\,km\,s^{-1}}$ 
 \citep[currently under discussion;][]{lallement2014}, 
total interstellar hydrogen density of $\mathrm{0.247\,cm^{-3}}$, 
improved ratios of radiation pressure over gravity $\beta$  for astronomical silicates. 
We calculated the grain-size dependent
variation of the  impact speed and impact direction using the dependence of radiation pressure 
upon particle size from \citet{kimura1999}, 
assuming that the grains are composed of astronomical silicates.

Our results confirm the existence of interstellar grains in the heliosphere in the 
size range from $\mathrm{0.05\,\mu m}$ to above $\mathrm{1\,\mu m}$. 
The overall size distribution measured in-situ with Ulysses within 5~AU from the Sun 
shows a deficiency of small grains below $\mathrm{0.3\,\mu m}$, compared to astronomically
observed interstellar dust in the interstellar medium \citep{mathis2000,draine2003,frisch2013a}. 
This deficiency can be partially explained by strong heliospheric filtering 
\citep{slavin2012,sterken2013a}. Up to now, no exact fit between the simulations and 
the data has proven this, but the general trend can be recognized.


We find a gas-to-dust mass ratio $R_{g/d} = 193^{+85}_{-57}$. This value is compatible with
gas-to-dust mass ratios 
derived from observations of sightlines to stars. Our analysis confirms earlier results 
 that 'big' (i.e.~$\mathrm{\approx 1\, \mu m}$-sized)
interstellar grains exist in the very local interstellar medium which 
are not easily accessible 
to astronomical observations \citep{wang2014}. 

\acknowledgments

We thank the Ulysses project at ESA and NASA/JPL for effective and successful mission operations,
and an anonymous referee for improving the presentation of our results.
HK is also grateful to Klaus Hornung for valuable discussions during the preparation of this 
manuscript. This research was supported by 
the German Bundesministerium f\"ur Bildung und Forschung through Deutsches
Zentrum f\"ur Luft- und Raumfahrt e.V. (DLR, grant 50\,QN\,9107). HK and PS 
gratefully acknowledge support by MPI f\"ur Sonnensystemforschung.
PS acknowledges support by Deutsche Forschungsgemeinschaft (DFG) grant KR 3621/1-1. 

 {\it Facilities:} \facility{Ulysses}.

\appendix

\section{Appendix}

Figure~\ref{fig_rotplot2} shows the Ulysses interstellar dust data set used in this paper for three
different grain size intervals with approximately equal numbers of particles in each figure.

\begin{figure}
\centering
\vspace{-6cm}
\hspace{-1.7cm}
\includegraphics[width=\textwidth]{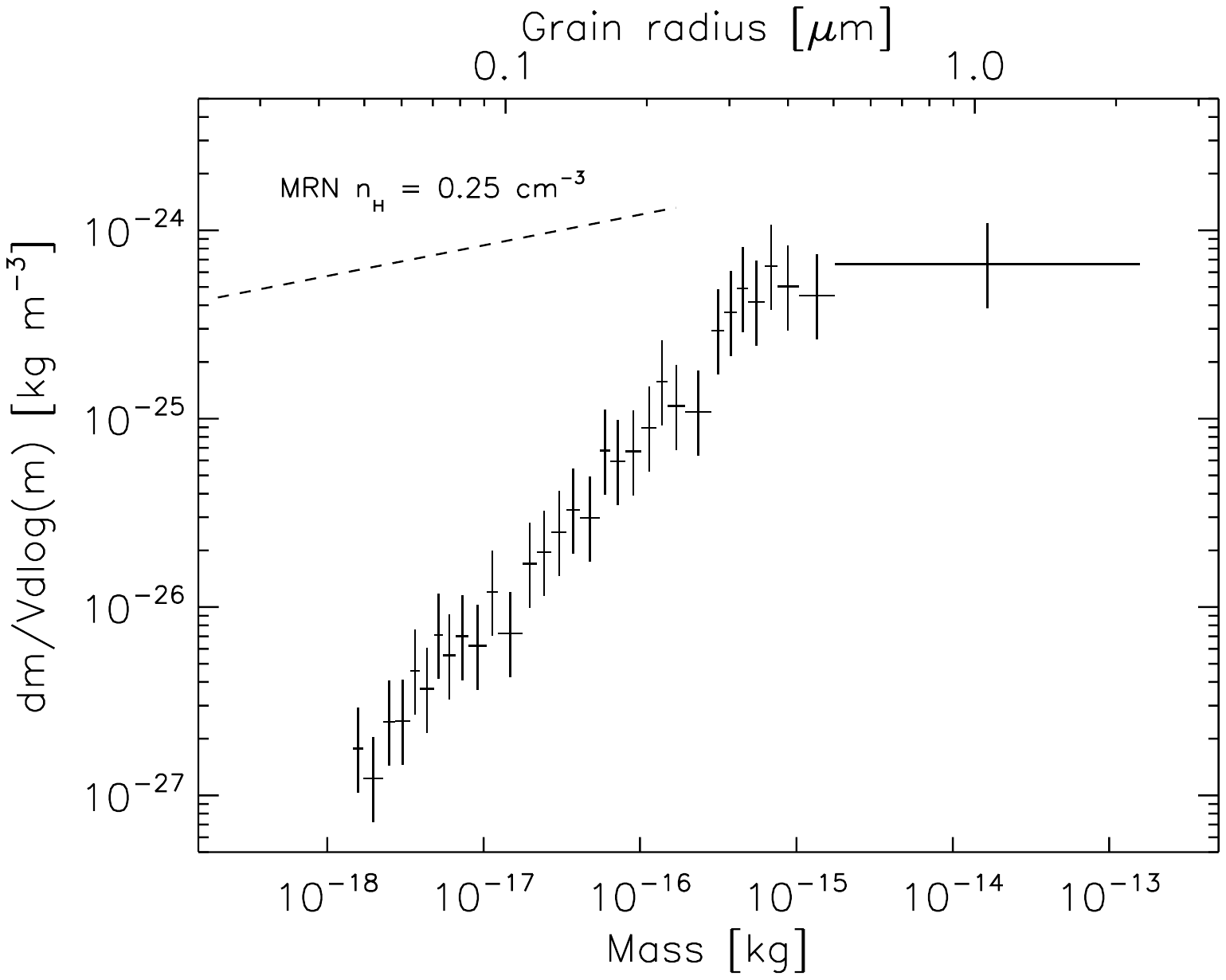}
\vspace{-7cm}
\caption[]{Mass distribution of interstellar grains derived from the Ulysses measurements
shown as mass per logarithmic mass interval 
and unit volume (987 particles). The approximate grain size for spherical particles with density 
$\rho = 3.3 \times 10^3\,\mathrm{kg\,m^{-3}}$ is 
shown at the top for comparison. The dashed line shows the mass distribution derived from 
astronomical observations \citep{mathis1977} for an interstellar hydrogen density of 
$0.25\,\mathrm{cm\,^{-3}}$. Grain masses were derived from the self-consistent model with 
accelerated ($\beta < 1$) and 
decelerated ($\beta > 1$) grains. The data are tabulated in Table~\ref{tab_massdist}.}
\label{fig_masshist}
\end{figure}

\begin{figure}
\vspace{-5cm}
\includegraphics[width=\textwidth]{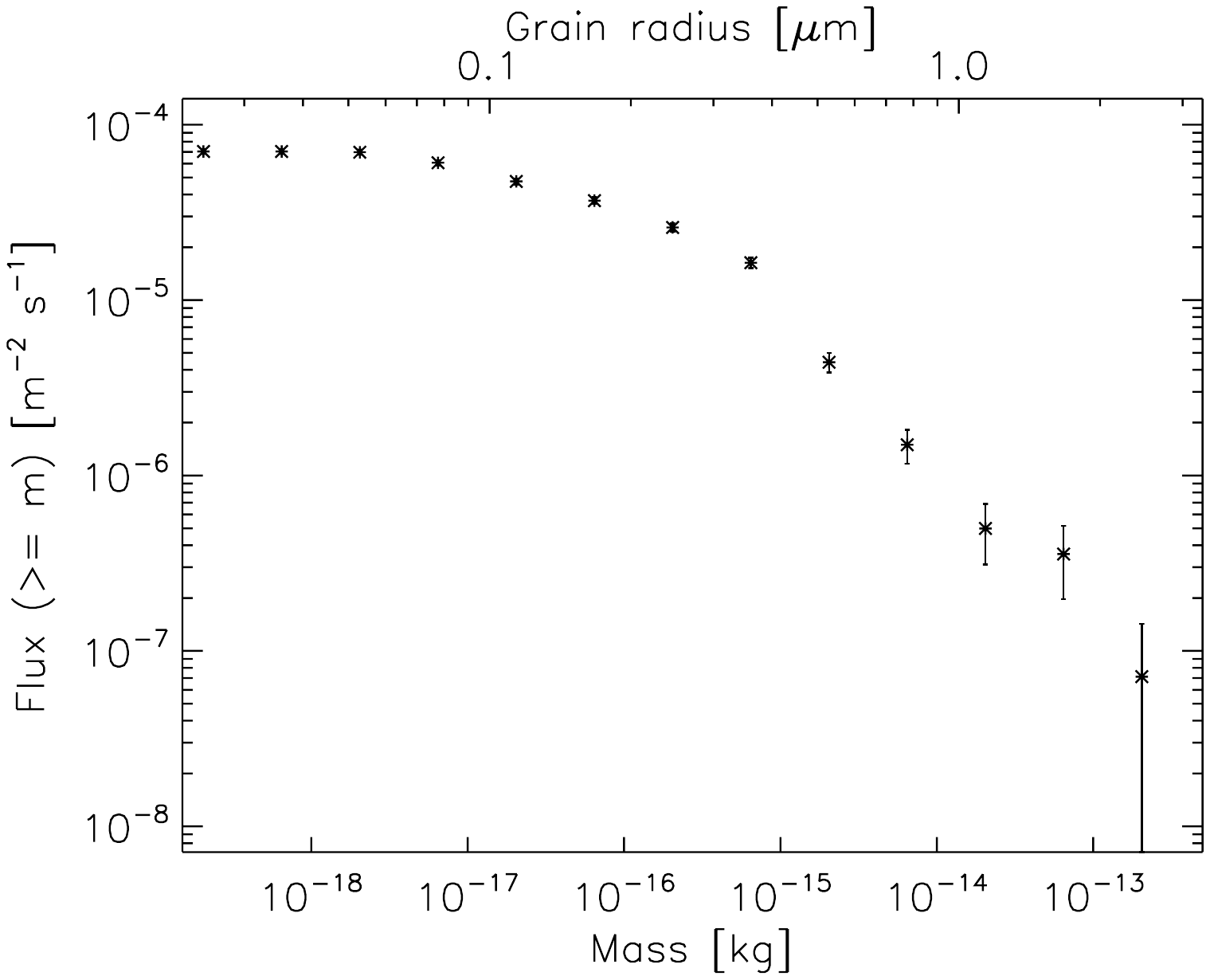}
\vspace{-7cm}
\caption[]{Flux of interstellar grains derived from the Ulysses measurements with the self-consistent model with 
accelerated ($\beta < 1$) and 
decelerated ($\beta > 1$) grains. The approximate grain size for spherical particles with density 
$\rho = 3.3 \times 10^3\,\mathrm{kg\,m^{-3}}$ is 
shown at the top for comparison. The data are tabulated in Table~\ref{tab_fluxdist}.}
\label{fig_flux}
\end{figure}

\begin{figure}
   \centering
   \parbox{15cm}{
   \vspace{-1.4cm}
\includegraphics[width=0.67\textwidth]{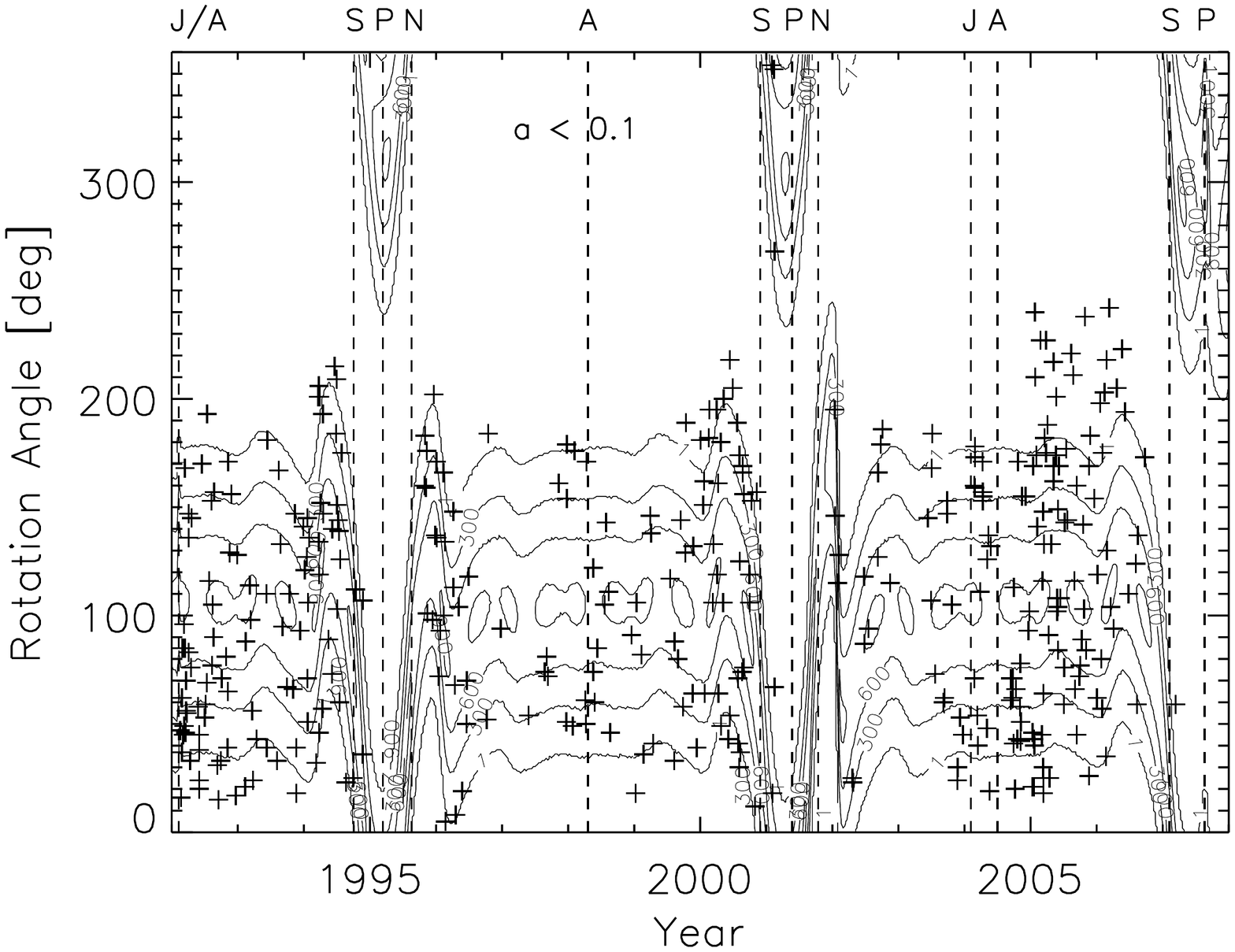}
\vspace{-1.9cm}
}
   \parbox{15cm}{
\includegraphics[width=0.67\textwidth]{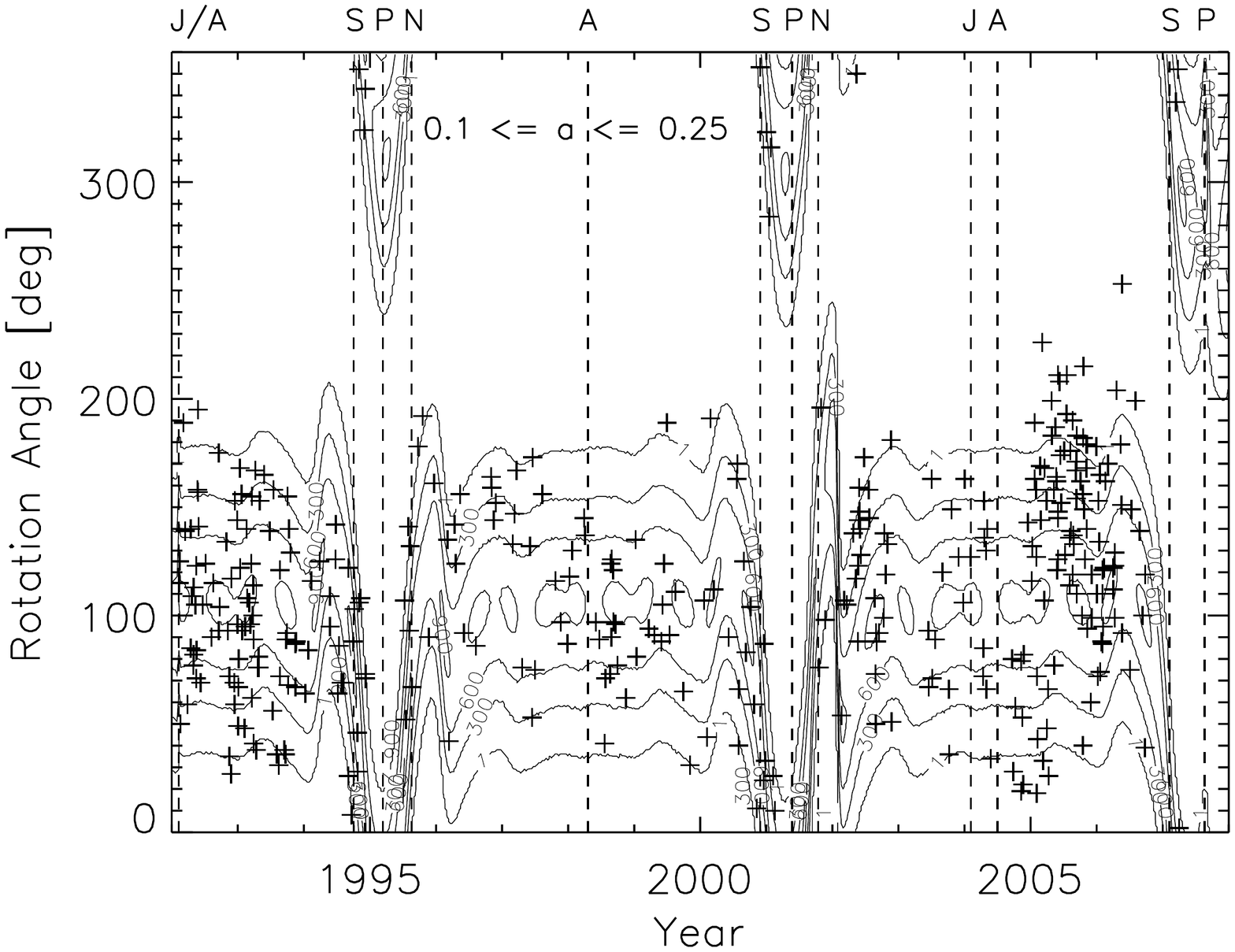}
\vspace{-0.7cm}
}
   \parbox{15cm}{
\vspace{-1.3cm}
\includegraphics[width=0.67\textwidth]{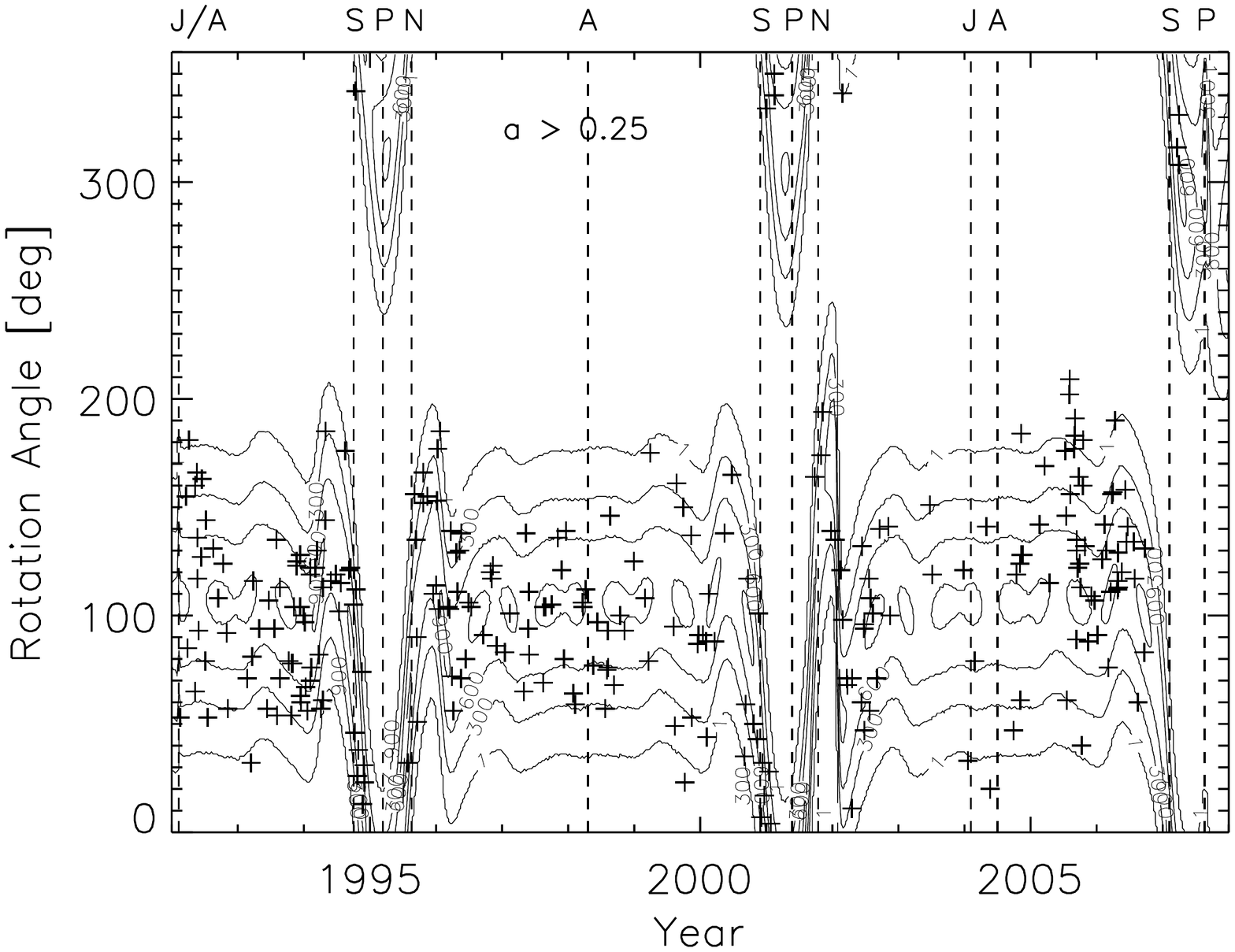}
\vspace{-0.9cm}
}
\caption[]{Same as Figure~\ref{fig_rotplot} but for different subsets of the interstellar dust data (assuming a grain
density of $\mathrm{3.3\,kg\,m^{-3}}$). Top: particles 
with radius $a < 0.1\,\mathrm{\mu m}$ (363 particles); Middle: $0.1 \,\mathrm{\mu m} \leq a \leq 0.25\,\mathrm{\mu m}$
(362 particles); Bottom: $a > 0.25\,\mathrm{\mu m}$ (262 particles).
}
\label{fig_rotplot2}
\end{figure}

The data shown in Figures~\ref{fig_masshist} and \ref{fig_flux} are listed in 
Tables~\ref{tab_massdist} to \ref{tab_fluxdist}.




\bibliography{pape,references}









\begin{table}
\caption[Charge signals and rise times measured by the dust detector]
{\label{tab_parameters} Parameters measured by the dust instrument upon 
impact of a dust particle onto the sensor and related parameters.
From \protect\citet{gruen1995a}. 
}
\small
\begin{tabular}{ll@{\hspace{4mm}}ccl}
                        &
                        &
                        &
                        &
                        \\[-2ex]
\hline
\hline
                        &
                        &
                        &
                        &
                        \\[-2ex]
Parameter/              &
\mc{1}{c}{Measured}     &
Range                   &
Accuracy                &
Related                 \\
digital value           &
\mc{1}{c}{quantity}     &
                        &
(logarithmic            &
particle                \\
                        &
                        &
                        &
steps)                  & 
parameters              \\[0.5ex]
\hline
                        &
                        &
                        &
                        &
                        \\[-1.7ex]
$    Q_E$/EA            & 
Negative charge         &
$\rm 10^{-14}-10^{-8}\,C$ &
48                      &
Mass, speed             \\
                        & 
(electrons)             &
                        &
                        &
                        \\
                        &
                        &
                        &
                        &
                        \\[-1.7ex]
$    Q_I$/IA            &
Positive charge         &
$\rm 10^{-14}-10^{-8}\,C$ &
48                      &
Mass, speed             \\
                        &
(ions)                  &
                        &
                        &
                        \\
                        &
                        &
                        &
                        &
                        \\[-1.7ex]
$    Q_C$/CA            &
Positive charge         &
$\rm 10^{-13}-10^{-9}\,C$ &
32                      &
Impact                  \\
                        &
(partially)             &
(channeltron            &
                        &
identification          \\
                        &
                        &
output)                 &
                        &
                        \\
                        &
                        &
                        &
                        &
                        \\[-3.0ex]
$    Q_P$/PA            &
Induced charge          &
                        &    
                        &
Electric charge         \\
                        &
\hspace{1ex} positive   &
$\rm 10^{-14}-10^{-12}\,C$ &
16                      &
                        \\
                        &
\hspace{1ex} negative   &
$\rm 10^{-14}-10^{-10}\,C$ &
32                      &
                        \\
                        &
                        &
                        &
                        &
                        \\[-1.7ex]
$    t_E$/ET            &
Rise time of            &
$\rm 10-100\,\mu s$     &    
16                      &
Speed                   \\
                        &
negative charge         &
                        &    
                        &
                        \\
                        &
                        &
                        &
                        &
                        \\[-1.7ex]
$    t_I$/IT            &
Rise time of            &
$\rm 10-100\,\mu s$     &
16                      &
Speed                   \\
                        &
positive charge         &
                        &
                        &
                        \\
                        &
                        &
                        &
                        &
                        \\[-1.7ex]
$    t_{EI}$/EIT        &
Time difference         &
$\rm -5-44\,\mu s$       &
16                      &
Impact                  \\
                        &
negative \& positive    &
                        &
                        &
identification          \\
                        &
charge signals          &
                        &
                        &
                        \\
                        &
                        &
                        &
                        &
                        \\[-1.7ex]
$    t_{PE}$/PET        &
Time difference         &
$\rm 1-400\,\mu s$      &
32                      &
Speed                   \\
                        &
induced \& negative     &
                        &
                        &
                        \\
                        &
charge signals          &
                        &
                        &
                        \\ [0.5ex]
\hline
\hline 
\end{tabular}
\end{table}


\begin{table}[th]
\caption{Criteria for the identification of interstellar dust grains used in this paper. \label{tab_criteria}}
\vskip4mm
\centering
\begin{tabular}{lll}
\hline 
Criteria                                  & Time Period/           & Comments \\
                                          & Spatial Region         & \\
\hline \hline
Rotation angle within $\pm 90^{\circ}$ of & Entire data set  & Sensor target plus side \\ 
interstellar helium flow direction        &                  & wall \\ \hline 
Rotation angle within $\pm 90^{\circ}$ of & 2005/2006        & Observed shift in rotation \\ 
interstellar helium flow direction &                  &  angle \\ 
plus  $40^{\circ}$ toward positive rotation                     &                      & \\ 
angles                                    &                  &          \\ \hline
$Q_I > \mathrm{10^{-13}\,C}$ for impacts at & Sun's polar regions & Removal of electromagne- \\
ecliptic latitude $|b| \geq 60^{\circ}$    &                     & tically accelerated grains \\ \hline
All dust impacts ignored with          & Inner solar system & No separation from inter- \\
$|b| < 60^{\circ}$ around perihelion &                    & planetary impactors possible \\ \hline
All dust impacts ignored in 39    & 1992/1993 and    & Jupiter dust streams  \\ 
 short time intervals defined     & 2002-2005        & removal \\ 
by \citet{baguhl1993a} and             &                     &         \\ 
\citet{krueger2006c} &            &         \\ \hline 
\end{tabular}
\end{table}

\begin{table}[h]
\caption{Mass distribution of interstellar grains derived in this paper. The data are shown in Figure~\ref{fig_masshist}. Column (1) 
lists the grain mass, column (2) the mass per logarithmic mass interval and unit volume, 
columns (3) and (4) give mass interval used for data binning (30 particles per mass bin), and columns (5) and (6) list the $\sqrt n$ 
error bars. \label{tab_massdist} }
\vskip4mm
\centering
\begin{tabular}{cccccc} \hline
  Mass  &   dm/Vdlog(m)       &  Err X+ &   Err X-  &   Err Y+  &   Err Y- \\
$\mathrm{[kg]}$&$\mathrm{[kg\,m^{-3}]}$ & $\mathrm{[kg\,m^{-3}]}$ & $\mathrm{[kg\,m^{-3}]}$ & $\mathrm{[kg\,m^{-3}]}$ & $\mathrm{[kg\,m^{-3}]}$ \\
  (1)   &    (2)              &   (3)   &    (4)    &    (5)    &    (6)   \\
 \hline
  1.67E-14   &   6.62E-25   &   1.57E-13   &   1.77E-15   &   1.10E-24   &   3.87E-25 \\
  1.35E-15   &   4.50E-25   &   1.77E-15   &   1.03E-15   &   7.44E-25   &   2.63E-25 \\
  8.84E-16   &   5.04E-25   &   1.03E-15   &   7.56E-16   &   8.34E-25   &   2.94E-25 \\
  6.87E-16   &   6.45E-25   &   7.56E-16   &   6.25E-16   &   1.07E-24   &   3.76E-25 \\
  5.55E-16   &   4.16E-25   &   6.25E-16   &   4.92E-16   &   6.88E-25   &   2.43E-25 \\
  4.53E-16   &   4.91E-25   &   4.92E-16   &   4.18E-16   &   8.13E-25   &   2.87E-25 \\
  3.81E-16   &   3.66E-25   &   4.18E-16   &   3.47E-16   &   6.07E-25   &   2.14E-25 \\
  3.15E-16   &   2.93E-25   &   3.47E-16   &   2.86E-16   &   4.85E-25   &   1.71E-25 \\
  2.36E-16   &   1.08E-25   &   2.86E-16   &   1.94E-16   &   1.80E-25   &   6.33E-26 \\
  1.71E-16   &   1.17E-25   &   1.94E-16   &   1.50E-16   &   1.94E-25   &   6.83E-26 \\
  1.38E-16   &   1.57E-25   &   1.50E-16   &   1.28E-16   &   2.61E-25   &   9.19E-26 \\
  1.14E-16   &   8.94E-26   &   1.28E-16   &   1.02E-16   &   1.48E-25   &   5.22E-26 \\
  9.04E-17   &   6.71E-26   &   1.02E-16   &   8.01E-17   &   1.11E-25   &   3.91E-26 \\
  7.19E-17   &   5.94E-26   &   8.01E-17   &   6.45E-17   &   9.83E-26   &   3.47E-26 \\
  5.97E-17   &   6.78E-26   &   6.45E-17   &   5.52E-17   &   1.12E-25   &   3.96E-26 \\
  4.78E-17   &   2.97E-26   &   5.52E-17   &   4.14E-17   &   4.92E-26   &   1.74E-26 \\
  3.74E-17   &   3.29E-26   &   4.14E-17   &   3.38E-17   &   5.44E-26   &   1.92E-26 \\
  3.03E-17   &   2.50E-26   &   3.38E-17   &   2.72E-17   &   4.14E-26   &   1.46E-26 \\
 \hline
 \end{tabular}
 \end{table}

\begin{table}[h]
\caption{Table~\ref{tab_massdist} continued. 
}
\vskip4mm
\centering
\begin{tabular}{cccccc} \hline
  Mass  &   dm/Vdlog(m)       &  Err X+ &   Err X-  &   Err Y+  &   Err Y- \\
$\mathrm{[kg]}$&$\mathrm{[kg\,m^{-3}]}$ & $\mathrm{[kg\,m^{-3}]}$ & $\mathrm{[kg\,m^{-3}]}$ & $\mathrm{[kg\,m^{-3}]}$ & $\mathrm{[kg\,m^{-3}]}$ \\
  (1)   &    (2)              &   (3)   &    (4)    &    (5)    &    (6)   \\
 \hline
  2.44E-17   &   1.96E-26   &   2.72E-17   &   2.18E-17   &   3.24E-26   &   1.14E-26 \\
  1.97E-17   &   1.70E-26   &   2.18E-17   &   1.77E-17   &   2.82E-26   &   9.93E-27 \\
  1.48E-17   &   7.24E-27   &   1.77E-17   &   1.23E-17   &   1.20E-26   &   4.22E-27 \\
  1.13E-17   &   1.20E-26   &   1.23E-17   &   1.04E-17   &   1.99E-26   &   7.03E-27 \\
  9.14E-18   &   6.22E-27   &   1.04E-17   &   8.01E-18   &   1.03E-26   &   3.63E-27 \\
  7.30E-18   &   7.01E-27   &   8.01E-18   &   6.65E-18   &   1.16E-26   &   4.09E-27 \\
  6.04E-18   &   5.55E-27   &   6.65E-18   &   5.48E-18   &   9.19E-27   &   3.24E-27 \\
  5.14E-18   &   7.11E-27   &   5.48E-18   &   4.82E-18   &   1.18E-26   &   4.15E-27 \\
  4.34E-18   &   3.68E-27   &   4.82E-18   &   3.90E-18   &   6.10E-27   &   2.15E-27 \\
  3.64E-18   &   4.58E-27   &   3.90E-18   &   3.39E-18   &   7.58E-27   &   2.67E-27 \\
  3.04E-18   &   2.49E-27   &   3.39E-18   &   2.72E-18   &   4.12E-27   &   1.45E-27 \\
  2.49E-18   &   2.46E-27   &   2.72E-18   &   2.28E-18   &   4.08E-27   &   1.44E-27 \\
  1.97E-18   &   1.23E-27   &   2.28E-18   &   1.71E-18   &   2.04E-27   &   7.18E-28 \\
  1.58E-18   &   1.77E-27   &   1.71E-18   &   1.46E-18   &   2.94E-27   &   1.04E-27 \\
  9.41E-19   &   1.95E-28   &   1.45E-18   &   6.12E-19   &   3.23E-28   &   1.14E-28 \\
 \hline
 \end{tabular}
 \end{table}

\clearpage

\begin{table}[h]
\caption{Cumulated flux distribution of interstellar grains derived in this paper. The data are shown in Figure~\ref{fig_flux}. \label{tab_fluxdist} Column (1) lists the grain mass, column (2) the cumulated flux of grains larger 
than the given mass, and columns (3) and (4) list the $\sqrt n$ errors.}
\vskip4mm
\centering
\begin{tabular}{cccc} \hline
  Mass  &   Flux ($\geq$m)    &   Err Y+  &   Err Y- \\
$\mathrm{[kg]}$&$\mathrm{[m^{-2}\,s^{-1}]}$ & $\mathrm{[m^{-2}\,s^{-1}]}$& $\mathrm{[m^{-2}\,s^{-1}]}$ \\
  (1)   &   (2)               &    (3)    &    (4)   \\
 \hline
  2.05E-19   &   7.03E-05   &   7.26E-05   &   6.81E-05 \\
  6.47E-19   &   7.03E-05   &   7.26E-05   &   6.81E-05 \\
  2.05E-18   &   6.97E-05   &   7.19E-05   &   6.75E-05 \\
  6.47E-18   &   6.06E-05   &   6.27E-05   &   5.86E-05 \\
  2.05E-17   &   4.75E-05   &   4.93E-05   &   4.56E-05 \\
  6.47E-17   &   3.69E-05   &   3.85E-05   &   3.53E-05 \\
  2.05E-16   &   2.59E-05   &   2.73E-05   &   2.46E-05 \\
  6.47E-16   &   1.63E-05   &   1.74E-05   &   1.52E-05 \\
  2.05E-15   &   4.42E-06   &   4.98E-06   &   3.86E-06 \\
  6.47E-15   &   1.50E-06   &   1.82E-06   &   1.17E-06 \\
  2.05E-14   &   4.99E-07   &   6.87E-07   &   3.10E-07 \\
  6.47E-14   &   3.56E-07   &   5.16E-07   &   1.97E-07 \\
  2.05E-13   &   7.13E-08   &   1.43E-07   &   0.00E+00 \\
 \hline
 \end{tabular}
 \end{table}

\clearpage


\clearpage

\clearpage

\clearpage

\clearpage

\end{document}